\documentclass[%
 reprint,
 amsmath,amssymb,
 aps,
]{revtex4-2}

\usepackage{graphicx}
\usepackage{dcolumn}
\usepackage{bm}
\usepackage{algorithm}
\usepackage{algpseudocode}

\begin{document}

\preprint{APS/123-QED}

\title{Geometric approach to nonequilibrium hasty shortcuts}

\author{Supraja S. Chittari}
\author{Zhiyue Lu}%
 \email{zhiyuelu@unc.edu}
\affiliation{Department of Chemistry, University of North Carolina at Chapel Hill, Chapel Hill, North Carolina 27599-3290, United States}%




\date{\today}

\begin{abstract}
Complex and even non-monotonic responses to external control can be found in many thermodynamic systems. In such systems, non-equilibrium shortcuts can rapidly drive the system from an initial state to a desired final state. One example is the Mpemba effect, where pre-heating a system allows a system to cool faster. We present nonequilibrium hasty shortcuts -- externally controlled temporal protocols that rapidly steer a system from an initial steady state to a desired final steady state. The term ``hasty'' indicates that the shortcut only involves fast dynamics without relying on slow relaxations. We provide a geometric analysis of such shortcuts in the space of probability distributions by using time-scale separation and eigenmode decomposition. We further identify the necessary and sufficient condition for the existence of non-equilibrium hasty shortcuts in an arbitrary system. The geometric analysis within the probability space sheds light on the possible features of a system that can lead to hasty shortcuts, which can be classified into different categories based on their temporal pattern. We also find that the Mpemba-effect-like shortcuts only constitute a small fraction of the diverse categories of hasty shortcuts. This theory is validated and illustrated numerically in the self-assembly model inspired by viral capsid assembly processes. 
\end{abstract}

\maketitle

\section{Introduction}
\label{sec:intro}
Systems driven out of equilibrium are prevalent across physics, biology, and material science. The Mpemba effect \cite{lu2017nonequilibrium,baity2019mpemba,klich2019mpemba,biswas2020mpemba,Gal2020-yq,degunther2022anomalous,yang2020non,bechhoefer2021fresh,carollo2021exponentially,yang2022mpemba,holtzman2022landau,PhysRevLett.119.148001,zhang2022theoretical,yang2020non,lin2022power,mpemba1969cool}, the strong Mpemba effect \cite{klich2019mpemba,biswas2020mpemba}, and certain hysteresis effects\cite{acharyya1998nonequilibrium} are general examples of systems demonstrating non-monotonic responses toward external control. Further, such complex nonequilibrium responses are also ubiquitous in natural systems from the cellular level, such as chemical reaction networks\cite{kobayashi2022hessian,rao2016nonequilibrium,qian2003stoichiometric,schmiedl2007stochastic} and circadian clocks\cite{zhang2020energy,weber2015entropy,wang2008potential}, to the molecular level, such as in DNA mechanics\cite{vtyurina2016hysteresis,baumann1997ionic,whitelam2008there,kapri2012hysteresis,doi:10.1073/pnas.071034098} or protein folding \cite{fasshauer2002snare,andrews2013hysteresis}. Timescale separation is common in such nonequilibrium processes, where dynamics of interest, such as collective behavior or self-assembly, occur on a much slower time scale than other components of system dynamics, such as vibrations of atoms due to thermal fluctuation. 

Traditional studies focus on the slow dynamics in systems, which is best suited for equilibrium in a static environment or slow relaxations in a slowly changing environment\cite{kurchan2005and}. In both cases, the fast dynamics can be integrated out and replaced by local Boltzmann distributions. Reconstructing the slow dynamics of complex systems has been the focus of recent efforts, such as using trajectory data mined from full atomistic simulations to generate few-state interpretable Markov state models.\cite{mardt2018vampnets, husic2018markov, harrigan2017msmbuilder} However, in the case of rapidly changing stimuli prevalent in realistic systems,\cite{fang2019nonequilibrium} both fast and slow dynamics play important roles in determining how systems evolve away from equilibrium \cite{parsons2017dimension}. Rapid environmental driving harnesses interesting fast dynamics and can result in a system being able to exist in a target far-from-equilibrium steady state\cite{horowitz2017minimum,PhysRevX.6.021022} or access nonequilibrium shortcuts as with the Mpemba effect\cite{lu2017nonequilibrium,baity2019mpemba,klich2019mpemba,biswas2020mpemba,Gal2020-yq,degunther2022anomalous,yang2020non,bechhoefer2021fresh,carollo2021exponentially,yang2022mpemba,holtzman2022landau,PhysRevLett.119.148001,zhang2022theoretical,yang2020non,lin2022power,mpemba1969cool}. Such far-from-equilibrium dynamics find a clean theoretical description in stochastic thermodynamics\cite{tome2015stochastic}. 

Understanding both the long-timescale macroscopic behavior and the fast relaxations at different environmentally controlled conditions is essential to elucidating how systems respond to rapidly changing external control protocols. In this paper, we present nonequilibrium hasty shortcuts -- temporal protocols comprising of rapidly switching sequences of control parameters $u$ -- that can steer a system from an initial equilibrium state (corresponding to the initial parameter value $u_{ini}$) to a final equilibrium state (corresponding to the final parameter value $u_{fin}$). The rapid parameter switches only allow the system to utilize its fast relaxations, but they do not leave excess time for slow relaxations. However, because the relaxation modes are not mutually perpendicular to each other (the evolution operator is not Hermitian or symmetric), both the fast and slow relaxation modes play important roles in understanding the fast relaxation dynamics of the hasty shortcuts. 
We developed a geometric approach to determine if a system can allow for a hasty shortcut by rapidly tuning a 1-dim control parameter. By writing down a family of master equations of a system at various values of the control parameter $u$'s and by performing a timescale separation analysis for each $u$, we obtain a family of fast projection operators to describe the fast evolution of the system at each transient control $u$, and a family of slow manifolds $\mathbb S(u)$ spanned by the slow relaxation eigenmodes. In the probability simplex comprising all possible probability distributions of the system's state, a fast projection operator for a given $u$ rapidly maps an arbitrary probability distribution onto the corresponding slow manifold. Based on the analysis of the family of fast projections and slow manifolds in the probability simplex, we developed a formalism towards identifying nonequilibrium hasty shortcuts through geometric analysis. This theory allows us to write down the necessary-and-sufficient condition for the existence of such nonequilibrium shortcuts in an arbitrary system that can be described by a master equation without degeneracy. 

To distinguish the hasty shortcut discussed in this paper from the many other interesting shortcuts defined in nonequilibrium thermal systems or nonequilibrium quantum systems, we provide a very brief review of several types of shortcuts. In quantum dynamics, the shortcuts to adiabaticity describe the problem where one can utilize an extra term of Hamiltonian to assist an evolving wave function in obeying adiabatic quantum evolution beyond the adiabatic limit (e.g, maintaining on the $n-$th eigenstate of a rapidly changing time-dependent Hamiltonian).\cite{jarzynski2013generating,chen2010fast,bason2012high} Similarly, in thermodynamics, there have been studies of shortcuts to isothermality,\cite{li2017shortcuts,li2022geodesic,gingrich2016near} where an extra term is engineered to keep a system in a Boltzmann distribution when environmental parameters rapidly change. There are several other types of shortcuts, or more precisely optimal control protocols, where one can engineer non-monotonic control of multiple environmental parameters to steer a system from one initial state to a final state with a minimum energy dissipation \cite{PhysRevLett.108.190602, rotskoff2015optimal} or waiting time\cite{Gal2020-yq,li2022cooling}.

In our work, we have defined a general class of hasty shortcuts-- by controlling an arbitrary control parameter (not limited to temperature) which rapidly switches in time, the system only relaxes the fast degrees of freedom, and it does not have enough time to relax its slow degrees of freedom due to the rapid control parameter switches. The hasty shortcut exists if the system, by only utilizing the series of fast relaxations corresponding to the sequence of control parameter values, can achieve the desired final equilibrium distribution. Some hasty shortcuts presented in this paper partially resemble the cooling/heating shortcuts proposed by the studies of the Mpemba effect, which claims that it may be faster to cool a system by first heating it up \cite{lu2017nonequilibrium,baity2019mpemba,klich2019mpemba,biswas2020mpemba,Gal2020-yq,degunther2022anomalous,yang2020non,bechhoefer2021fresh,carollo2021exponentially,yang2022mpemba,holtzman2022landau,PhysRevLett.119.148001,zhang2022theoretical,yang2020non,lin2022power,mpemba1969cool}. It has been shown that heating and cooling are inherently asymmetric \cite{ibanez2023heating}, and thus interesting shortcuts may emerge by non-monotonically controlling temperature.  
Some of the hasty shortcuts we found in this work resemble the strong Mpemba effect: to drive a system from $u_{ini}$ to $u_{fin}>u_{ini}$, the hasty shortcut may start by an initial downward switch of $u_1<u_{ini}$, which is followed by a final quench of $u_2=u_{fin}$. 

To illustrate the geometric theory of hasty shortcuts, we demonstrate a simple assisted-assembly model inspired by a viral capsid assembly problem studied by Phillip Geissler's group \cite{rotskoff2018robust}. We provide a simple 24-state assembly model that demonstrates timescale separation. In this rather kinetically simple model with a single control parameter (the subunit concentration $c_s$), we demonstrate numerous hasty shortcuts connecting different equilibria.

\section{Theory}
\label{sec:Theory}
\subsection{Timescale separation at constant control}
\label{subsec:time_sep}
We describe a system's dynamics by a time-dependent master equation, 
\begin{equation}
\label{eq:master}
\frac{d \vec p(t)}{dt} = \hat R\big(u(t)\big) \cdot \vec p(t)
\end{equation}
where $\vec p(t)$ is a probability vector describing $n$ configurations and $\hat R(u)$ is the transition rate matrix controlled by a parameter $u$. Here the control parameter is allowed to vary in time according to a temporal protocol $u(t)$. 

If the control parameter is held fixed at a constant value $u$, then regardless of the initial probability distribution $\vec p(0)$, the system eventually relaxes towards its corresponding stationary state $\vec p^{~ss}_u$, where $\hat R(u) \vec p^{~ss}_u = 0$. Such relaxation within a stationary environment (i.e., constant control parameter $u$) can be separated into different relaxation eigenmodes, where each mode is an eigenvector of $\hat R(u)$:
\begin{equation}
    \hat R(u) \vec v_i (u) = \lambda_i(u) \vec v_i (u)
\end{equation}
and the corresponding eigenvalue $\lambda_i(u)$ dictates the relaxation rate. The time evolution of a system's probability distribution while the control parameter is maintained at a stationary value $u$ is 
\begin{equation}
\label{eq:solution}
    \vec p(t) = \sum_i  b_i \vec v_i(u) e^{\lambda_i(u) t}
\end{equation}
Notice that an initial probability vector $\vec p(0)= \sum_i  b_i \vec v_i(u)$ is a superposition of all of the eigenvectors with weighting factor $b_i$'s. Here we assume that the rate matrix $\hat R(u)$ is diagonalizable for each $u$ involved (e.g., this assumption is guaranteed to be true when detailed balance condition is satisfied).
In another description, one can define an evolution operator for a constant environmental condition $u$ over time period $t$, which maps an initial probability vector $\vec p(0)$ into $\vec p(t)$, the probability vector at time $t$. This evolution map can be written as
\begin{equation}
\label{eq:solution1}
    \vec p(t) = \hat V e^{\hat \Lambda t} [\hat V]^{-1} \vec p(0) = \hat V \hat D_t [\hat V]^{-1}  \vec p(0) = \hat W_t ~\vec p(0)
\end{equation}
where $\hat V$ is a square matrix consisting of all of the eigenvectors as its columns, and the $\vec b= [\hat V]^{-1}  \vec p(0)$ defines the weighting factor $ b_i$ that decomposes the initial probability vector into a superposition of eigenvectors. Here we have defined a propagation operator
\begin{equation}
    \hat W_t = \hat V \hat D_t [\hat V]^{-1}
\end{equation}
where $\hat D_t$ is a diagonal matrix, with its diagonal elements represented by
\begin{equation}
    \hat D_{t,ii}={\rm exp}(\lambda_i  t)
\end{equation}

\begin{figure}[h]
    \includegraphics{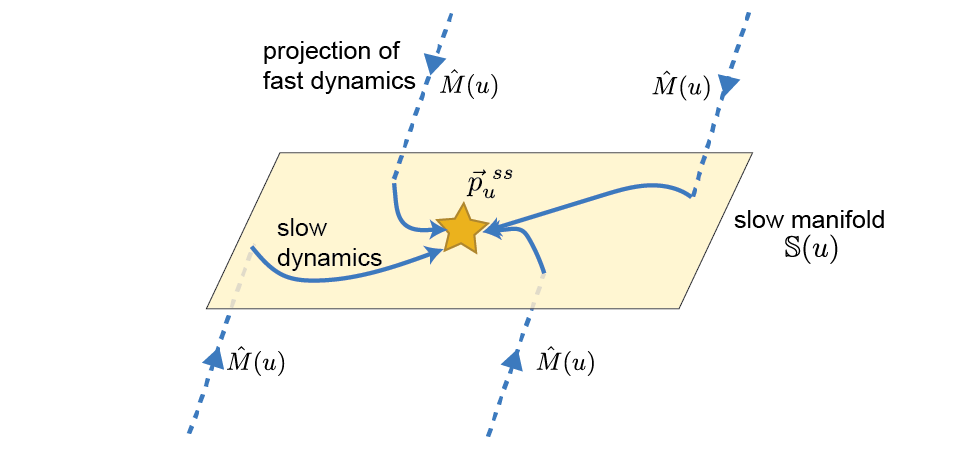}
    \caption{Schematic of separation of fast dynamics and slow dynamics in the space of probability distributions (probability simplex). Each point in the space represents a probability distribution, and the golden star represents the stationary distribution for a given control $u$, $\vec p^{~ss}_u$. The slow manifold $\mathbb S(u)$, shown by the yellow plane is spanned by all the slow eigenvectors. The fast relaxation described by the rapid projection operator $\hat{M}(u)$ maps any initial probability distributions rapidly to a point on the slow manifold. The fast projections form a set of parallel dashed lines that are not necessarily perpendicular to the slow manifold. After a rapid projection, if a system is allowed an infinitely long time to fully relax at the constant control $u$, it traverses toward the final stationary state $\vec p^{~ss}_u$ via the solid blue arrows.}
    \label{fig:dimension_reduction}
\end{figure}

In our analysis, the eigenvalues are arranged in descending order, with the largest eigenvalue always being 0 and every other eigenvalues' real components being negative \cite{friedland2013perron}. 
In this paper, we are interested in systems that have separated relaxation timescales. In this system, for a transiently stationary control parameter value $u$, the relaxation dynamics can be separated into fast and slow eigenmodes. Reflected in the spectrum of eigenvalues, if
\begin{equation}
\label{eq:gap}
    \frac{1}{\lambda_1},\frac{1}{ \lambda_2}, \cdots ,\frac{1}{\lambda_c},\frac{1}{ \lambda_{c+1}} \cdots, \frac{1}{ \lambda_n}
\end{equation}
have a gap between $1/\lambda_c$ and $1/\lambda_{c+1}$, then the separation of timescale occurs.
The timescale separation allows us to define the persistence time $\tau$, which is infinitely short for the slow relaxations and infinitely long for the fast relaxation:
\begin{equation}
\label{eq:tau}
    \frac{1}{|\lambda_{c+1}|} \ll \tau \ll \frac{1}{|\lambda_c|}
\end{equation}
At this timescale of $\tau$, $e^{\lambda_i \tau} \approx 0$ if $i > c$ and $e^{\lambda_i \tau} \approx 1$ if $i \leq c$. 
In other words, the constant $u$-dynamics of time duration $\tau$ at fixed control value $u$, denoted by $\hat W_\tau$, can be approximated by a rapid projection operator $\hat M(u)$, which maps the full probability space to the slow manifold $\mathbb S(u)$ (see Fig.~\ref{fig:dimension_reduction})
\begin{equation}
\label{eq:WMQ}
    \hat W_\tau \approx \hat M(u) =\hat V \hat D^{(01)} [\hat V]^{-1}
\end{equation}
where $\hat D^{(01)}$ is a diagonal matrix populated by $0$'s and $1$'s. Further, the first $c$ entries equal $1$ (i.e., $\hat D^{(01)}_{ii}=1$ for $i\leq c$) as $\tau$ is too short to allow for any slow relaxation, and $\hat D^{(01)}_{ii}=0$ for $i > c$ as within the time $\tau$, all the fast eigenmodes fully relax and vanish. This map $\hat M(u)$, defined for a control parameter $u$, is a projection operator that characterizes the rapid relaxation of any initial probability to a slow manifold $\mathbb S(u)$. The slow manifold is a vector space spanned by the eigenvectors corresponding to the slow modes. At the timescale $\tau$, the slow relaxations do not have enough time to occur.

If one further waits for a time much longer than $\tau$, the slow dynamics defined within the slow manifold can eventually bring the system to the ultimate stationary state, $\vec p^{~ss}_u$. As illustrated in Fig.~\ref{fig:dimension_reduction}, at a given stationary condition $u$, the fast projection onto the slow manifold $\hat M(u)$ are represented by parallel bundles of dashed lines (fibers) onto the slow manifold $\mathbb S(u)$(a yellow hyperplane); given long enough time $t\gg \tau$, the slow relaxations toward the ultimate stationary state $\vec p^{~ss}_u$ are shown by the solid arrows confined on the slow manifold $\mathbb S(u)$.

\subsection{Nonequilibrium hasty shortcuts}
\label{subsec:hasty_shortcut}

We define the nonequilibrium hasty shortcut as a temporal protocol which is a sequence of rapidly switched control values $u_{ini}, u_1, u_2,\cdots, u_{fin}$ that can steer the system from the initial stationary state $\vec p^{~ss}_{u_{ini}}$ to the final stationary state $\vec p^{~ss}_{u_{fin}}$. The term ``hasty'' indicates that the dwell time $\tau$ at each control value $u_i$ is infinitely short to allow for the slow relaxations to occur (see Eq.~\ref{eq:tau}). Since the controlled protocol avoids waiting for slow relaxations that are exponentially slower than fast relaxations, the protocol serves as a shortcut. In summary, the whole control protocol constructs an evolution operator comprising of a sequence of fast relaxation projections $\hat M(u)$'s, which rapidly maps the initial state to the final state:
\begin{equation}
\label{eq:defshortcut}
    \vec p^{~ss}_{u_{fin}} = \hat M(u_{fin}) \cdots \hat M(u_2) \hat M(u_1) \vec p^{~ss}_{u_{ini}}
\end{equation}

\begin{figure}[h]
    \includegraphics{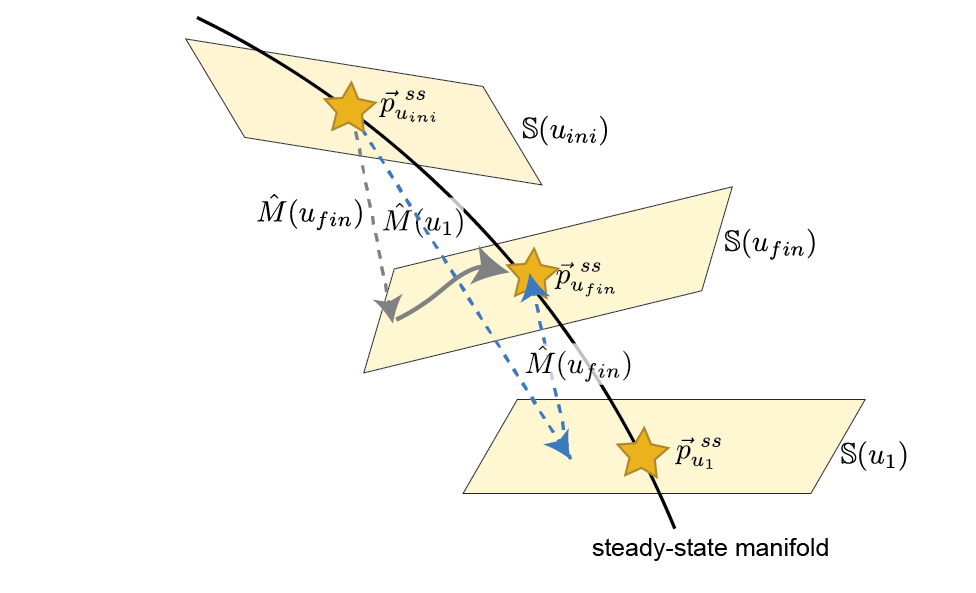}
    \caption{Schematic comparison between a hasty shortcut and a rudimentary single-step quench control to drive a system from an initial steady state $\vec p^{~ss}_{u_{ini}}$ to a desired final steady state $\vec p^{~ss}_{u_{fin}}$. The hasty shortcut utilizes an intermediate control parameter value $u_1$. The fast projections $\hat M(u_1)$ and $\hat M(u_{fin})$ are shown as two dashed blue lines that spend a short time at each value of $u$. In the shortcut, the system achieves the desired final state without waiting for any slow relaxations. In comparison, the rudimentary single-step quench takes a long time with an initial fast relaxation (dashed gray arrow) followed by a slow relaxation (solid gray arrow) on the slow manifold $\mathbb S(u_{fin})$.}
    \label{fig:shortcuts}
\end{figure}

Let us illustrate the nonequilibrium hasty shortcut by comparing it with a straightforward rudimentary control protocol that directly sets the control value from $u_{ini}$ to $u_{fin}$. In Fig.~\ref{fig:shortcuts}, we sketch three slow manifolds $\mathbb S(u_{ini})$, $\mathbb S(u_{1})$, and $\mathbb S(u_{fin})$ within the probability simplex. The black locus consists of all stationary probability distributions $\vec p^{~ss}_{u}$ for arbitrary values of $u$. The control task is to steer the system from the initial steady state $\vec p^{~ss}_{u_{ini}}$ to the final steady state $\vec p^{~ss}_{u_{fin}}$. 
A hasty shortcut is illustrated by a control sequence ($u_{ini},~u_1,~u_{fin}$). During the first step of the shortcut, the system rapidly evolves to the slow manifold $\mathbb S(u_1)$ by the rapid projection $\hat M(u_1)$. Then without waiting for any slow relaxation, the second step control (of value $u_{fin}$) directly evolves the system into the desired final steady state by the rapid projection $\hat M(u_{fin})$.
In comparison, the rudimentary 1-step quenching control, where $u_{ini}$ is directly set to $u_{fin}$ involves both an initial rapid projection $\hat M(u_{fin})$ that maps the system onto the slow manifold $\mathbb S(u_{fin})$ and a long-time slow relaxation as the system slowly traverses the slow manifold toward the final state $\vec p^{~ss}_{u_{fin}}$. In summary, the hasty shortcut can save time by avoiding slow relaxations, which take exponentially longer time than rapid relaxations. 

The nonequilibrium hasty shortcut can be analogous to a hasty driver who is averse to any traffic slowdowns. By constantly resetting the destination on the car's GPS, the driver aims to always travel at high speed until ultimately colliding with the final destination at full speed. In this analogy, traveling at high speed corresponds to utilizing fast relaxations, traffic slowdowns represent slow dynamics on the slow manifold, and transiently resetting the destination of GPS corresponds to the rapid switches of $u$. Notice that some hasty shortcuts, e.g., the one shown in Fig.~\ref{fig:shortcuts}, may be counter-intuitive and resemble the Mpemba effect, as it initially steers $u$ away from the direction of the target $u_{fin}$ to reach an intermediate state, which can then be rapidly relaxed to the desired final state by $\hat M(u_{fin})$. In Sec.~\ref{subsec:num_hasty}, we classify hasty shortcuts into different categories and show that the Mpemba-effect-like shortcuts only comprise a small fraction of possible hasty shortcuts. 

\subsection{Condition for the existence of nonequilibrium hasty shortcuts}
\label{subsec:geo_exist} 
We examine the condition of a system to allow for the nonequilibrium hasty shortcut. The hasty shortcut presented in this paper is likely to be found in kinetically nontrivial systems, including anti-ferromagnetic spin systems where one can easily identify the strong Mpemba effect\cite{klich2019mpemba,Gal2020-yq}. However, the hasty shortcuts are more general than the Mpemba effect and thus could be found in systems where the Mpemba effect does not exist. Here we use a geometric approach to identify the general kinetic features that allow for a general thermal system to have nonequilibrium hasty shortcuts. 

Let us first identify the necessary and sufficient condition for the existence of non-equilibrium hasty shortcuts. Without loss of generality, consider a system described by a family of master equations parametrized by a one-dimensional control variable $u$. We further assume that the rate matrix $\hat R(u)$ is diagonalizable for each $u$ involved in our control. The rate matrix $\hat R(u)$ of ergodic systems satisfying detailed balance conditions and the rate matrix $\hat R(u)$ of non-detail-balanced systems without degenerate eigenvectors both guarantee diagonalizablility. For these systems, the steady states $\vec p^{~ss}_u$ for arbitrary given values of $u$ form a 1-dimensional locus $L$:
\begin{equation}
\label{eq:Locus}
    L= \bigcup_{u} \vec p^{~ss}_u
\end{equation}
which we will refer to as the ``steady-state manifold''. If, for a system, any stationary distribution $\vec p\in L$ can be mapped to a different stationary distribution in $L$ by a sequence of rapid projection operators, $\hat M(u)$'s, then the system allows for hasty shortcuts. 


\begin{figure}[h]
    \includegraphics{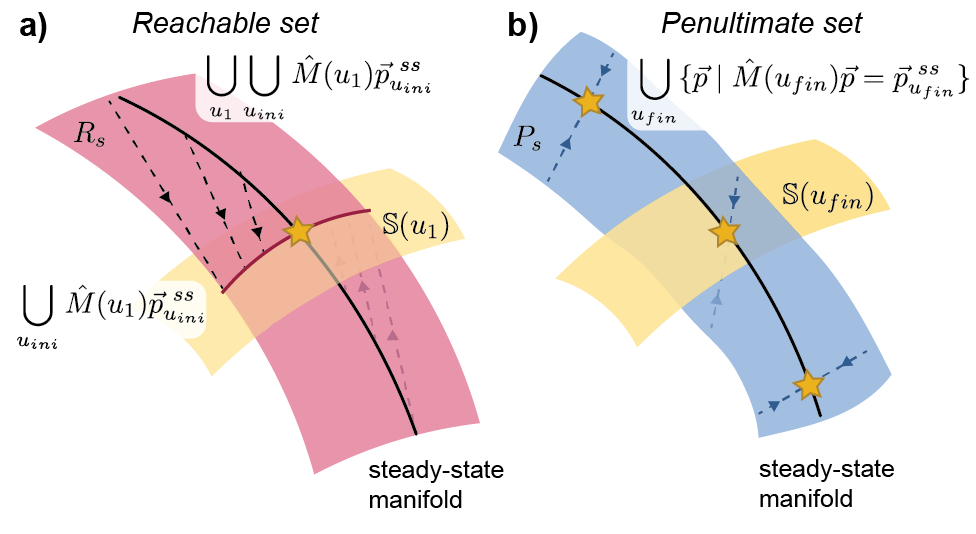}
    \caption{Illustration of reachable and penultimate sets in the $\vec p$-space. a) The single-step reachable set $R_s(1)$ (illustrated as a red ribbon) is generated as the union of all states reachable by the projection of any arbitrary initial steady-state $\vec p^{~ss}_{u_{ini}}\in L$ (black curve) to any slow manifold $\mathbb R(u_1)$. Given one value of $u_1$, a transient set $T(u_1)$ (red curve) within the slow manifold $\mathbb S(u_1)$ (yellow ribbon) is the projection of $L$ (black curve) by $\hat M(u_1)$ (dashed arrows). The reachable set $R_s(1)$ is the red ribbon swept by the red curve $T(u_1)$ for all choices of $u_1$. b) The penultimate set $P_s$ (illustrated as a blue ribbon) is the set of all $\vec p$'s that can evolve via an ultimate-step projection $\hat M(u_{fin})$ to reach the corresponding steady-state distribution $\vec p^{~ss}_{u_{fin}}$ for all possible values of $u_{fin}$. The ultimate projections corresponding to three choices of $u_{fin}$ are sketched as three dashed blue arrows toward three golden stars. c) Intersection of the transient set $T(u_1)$ (red solid line) with a fiber $F(u_1)$ (blue dashed line) on the slow manifold hyperplane (yellow ribbon) occurs at the steady state $p^{~ss}_{u_1}$.}
    \label{fig:departure_arrival}
\end{figure}

The necessary and sufficient condition for hasty shortcuts can be geometrically represented by the nontrivial intersections between two sets constructed below, where both sets can be illustrated within a system's probability simplex (or the $\vec p$-space).

\textbf{Reachable Set}, $R_s(n)$, is defined as the set of probability distributions $\vec p$ that are reachable through steering any initial stationary distribution $\vec p^{~ss}_{u_{ini}}$ by arbitrary $n$-step sequences of fast relaxation projections $\hat M (u_i)$'s: 
\begin{equation}
\label{eq:Rs}
    R_s(n)=\bigcup_{u_{n}} \cdots \bigcup_{u_{1}} \bigcup_{u_{ini}} \hat M(u_{n}) \cdots \hat M(u_1) \vec p_{u_{ini}} 
\end{equation}
The $n$-step reachable set $R_s(n)$ can be constructed iteratively by 
\begin{equation}
\label{eq:Rs_iter}
    R_s(i+1)=\bigcup_{u} \hat M(u) R_s(i)
\end{equation}
Since $\hat M$ is a projection operator, $R_s(i)\subset R_s(i+1)$ is true for all positive integer $i$. Additionally, since the $\vec p$ is bounded, we can denote the bounded infinite-step reachable set as $R_s = \lim_{n\rightarrow \infty} R_s(n)$. For illustration, in Fig.~\ref{fig:departure_arrival}(a), we sketch the construction of $R_s(1)$ in the $\vec p$-space. In this case, the set $R_s(1)$ consists of all possible distributions $\vec p$'s that can be reached by any single-step rapid projection $\hat M(u_1)$ applied to any initial steady-state distribution $\vec p^{~ss}_{u_{ini}}$, for any arbitrary $u_1$ and $u_{ini}$. This reachable set is illustrated by a 2-dim ribbon, which is swept by a moving 1-dim curve ($T(u)$) in the $\vec p$-space: given a chosen value of $u_1$, the rapid projection $\hat M(u_1)$ maps all possible steady states in $L$ to a transient set 
\begin{equation}
\label{eq:Tu1}
    T(u_1)\equiv \hat M(u_1) L = \bigcup_{u_{ini}}\hat M(u_1)p^{~ss}_{u_{ini}}
\end{equation}
which is a 1-dim curve within the slow manifold $T(u_1) \in \mathbb S(u_1)$. In Fig.~\ref{fig:departure_arrival}a, the transient set $T(u_1)$ is the red curve obtained by the projection of the whole steady-state manifold $L$ (as black curve) onto the slow manifold $\mathbb S(u_1)$ (as a yellow plane) via the single-step projection $\hat M(u_1)$ (as dashed arrows). Then one can construct the set $R_s(1)$ as the union of all transient sets $T(u_1)$ for all possible values for $u_1$:
\begin{equation}
\label{eq:Rs1}
    R_s(1)=\bigcup_{u_1}T(u_1) 
\end{equation} 
Geometrically, in Fig.~\ref{fig:departure_arrival}a, the set $R_s(1)$ is the pink ribbon sweep by the red curves for different choices of $u_1$. It is straightforward to see that $L$, the set of steady states (i.e., the black curve), is a backbone of $R_s(n)$ for all positive integer $n$.


\textbf{Penultimate Set} is defined as 
\begin{equation}
\label{eq:Ps}
  P_s=\bigcup_{u}~ F(u)= \bigcup_{u} \{ \vec p \mid \hat M(u) \vec p = \vec p^{~ss}_{u} \}
\end{equation}
as the set of all possible $\vec p$'s that can be directly projected to $\vec p^{~ss}_u$ by a single-step rapid relaxation $\hat M(u)$. Here we have defined a ``fiber'' 
\begin{equation}
\label{eq:Fiber}
    F(u) =  \{ \vec p \mid \hat M(u) \vec p = \vec p^{~ss}_{u} \}
\end{equation}
as a multi-dimensional set of distributions that can be immediately relaxed into $\vec p^{~ss}_{u}$ by the rapid projection $\hat M(u)$.
Notice that the fiber $F(u)$ is not necessarily a 1-dim object, and the number of fast eigenmodes for $\hat R(u)$ determines $F(u)$'s dimension.
In Fig.~\ref{fig:departure_arrival}b, each dashed arrow represents one fiber $F(u)$ for a specific value of $u$, and the penultimate set can be considered as a multi-dimension blue ribbon swept by a family of fiber $F(u)$ for all values of $u$.
One can show that the steady-state manifold $L$ is a subset of the penultimate set. In other words, the set of steady states $L$ (as black curve) is the backbone of the penultimate set.

\textbf{Existence condition of hasty shortcuts:} The existence condition of (up-to-n-step) hasty shortcuts can be written as the non-trivial intersection between the $n$-step reachable set $R_s(n)$ and the penultimate set $P_s$:
\begin{equation}
\label{eq:condition}
    R_s(n)\cap P_s ~~\supsetneq ~L
\end{equation}
where the not-equal-superset sign, $\supsetneq$, is used to exclude the trivial intersection between $R_s(n)$ and $P_s$. Here trivial intersection between $R_s(n)$ and $P_s$ is their shared backbone $L$ (i.e., the set of steady-state distributions). Intuitively, the trivial intersections between the $R_s(n)$ and $P_s$ correspond to time-independent protocol $u_{fin}=u_n=\cdots=u_1=u_{ini}$, where the initial and final steady-state distributions are trivially identical. In the following, we discuss the geometrical features that allow for the non-trivial intersection. 


\subsection{Geometric analysis}
\label{subsec:geo_eigen}
The geometric representation of the existence condition for hasty shortcuts (as Eqs.~\ref{eq:Locus},~\ref{eq:Rs},~\ref{eq:Ps}, and \ref{eq:condition}) allows us to investigate the underlying mechanisms of hasty shortcuts.  

Firstly, we show that the hasty shortcuts are inherently far from equilibrium and steady states, which allows for the necessary geometrical features for the non-trivial intersection between $R_s(n)$ and $P_s$. This can be visualized by Fig.~\ref{fig:departure_arrival}c: recall that $R_s(1)=\cup_u T(u)$ is the union of a family of transient sets $T(u)$ for all $u$'s, and $P_s=\cup_u F(u)$ is the union of a family of ``fibers'' $F(u)$ for all $u$'s. Also, recall that the existence condition for hasty shortcuts is that $R_s$ and $P_s$ intersect beyond the shared steady-state backbone $L$ (Eq.~\ref{eq:condition}). However, as shown in Fig.~\ref{fig:departure_arrival}c, at a given value $u$, $T(u)$ and $F(u)$ can only intersect at the corresponding steady state, $\vec p^{~ss}_u$. This is because $F(u)$ belongs to the subspace spanned by fast eigenvectors, and $T(u) \subset \mathbb S(u)$ is confined to the slow manifold spanned by the slow eigenvectors. Thus, we argue that for systems $\hat R(u)$'s with continuous dependence on control parameter $u$, in the vicinity of $u$, it is very unlikely to construct non-trivial intersections between the reachable set and the penultimate set. In other words, hasty shortcuts can only be achieved by far-from-steady-state driving protocols.

Further, we illustrate potential geometric features that could allow for the existence of hasty shortcuts. We propose a few geometric features that allow for the non-trivial intersection between $R_s$ and $P_s$. They shed light on the kinetic features that could allow a system to have hasty shortcuts. Even though $T(u)$ and $F(u)$ cannot intersect beyond the trivial point $\vec p^{~ss}_u$ (see Fig.~\ref{fig:departure_arrival}c), the non-trivial intersection between $R_s(n)$ and $P_s$ may still occur if any one or more of the following geometrical features are satisfied:
\begin{itemize}
    \item $(\#1)$ the steady-state locus is a highly bent curve that almost intersects itself;
    \item $(\#2)$ the reachable set and/or the penultimate set are highly bent;
    \item $(\#3)$ the reachable set and/or the penultimate set occupies a large space in the probability simplex.
\end{itemize}  

{\boldsymbol{$(\#1)$} } states that the curve $L$ is so highly curved that it almost approaches self-intersection, causing some point $\vec p^{~ss}_{u_a}$ along the curve to be in close proximity to another point $\vec p^{~ss}_{u_b}$ at a very different $u$. In this situation, the shared backbone of the two ``ribbons'' $R_s$ and $P_s$ is highly bent and almost approaches self-intersection, increasing the chance for non-trivial intersections between $R_s$ and $P_s$. To the next level, if the curve $L$ is so bent that its point $p^{~ss}_{u_a}$ intersects the fiber $F(u_b)$ corresponding to $\hat M (u_b)$, it indicates that there exists a single-step quench shortcut from $p^{~ss}_{u_a}$ to $p^{~ss}_{u_b}$, which can be realized by directly quenching $u_a$ to $u_b$. In the extreme case, if the curve $L$ intersects itself at $p^{~ss}_{u_a}=p^{~ss}_{u_b}$, the system may demonstrate reentrant behaviors that resemble those studied in reentrant transitions\cite{milin2018reentrant, banerjee2017reentrant,mitchell2021real,redner2013reentrant,doi:10.1119/1.18681,camerin2020microgels}. It is worth pointing out that the quenching hasty shortcut due to the almost self-intersecting curve $L$ has been demonstrated in mean field anti-ferromagnetic Ising model that demonstrates Mpemba effects \cite{Gal2020-yq,klich2019mpemba}. For other general thermodynamic systems controlled by a simple control parameter $u$ (such as temperature, pressure, or concentration), the feature $(\#1)$ may not be easily satisfied. 
For general systems without a strongly non-monotonic parameter dependence required in geometric feature $(\#1)$, the hasty shortcut may still exist under geometric features $(\#2)$ and $(\#3)$. 

{\boldsymbol{$(\#2)$} }  states that the geometric shape of either $R_s(n)$ or $P_s$ to be highly bent. It increases the chance for the two sets to intersect beyond their shared backbone $L$, which is the condition of the existence of the hasty shortcut. Notice that here we do not assume for geometric feature $(\#1)$, and thus the contortion of $R_s$ or $P_s$ is not due to a bent backbone. Rather, the contortion of $R_s$ and/or $P_s$ can result from the change of fast and slow eigenvectors $\vec v_i(u)$ of the rate matrix $\hat R(u)$ as $u$ varies. In other words, even for systems whose steady states $\vec p^{~ss}_u$ have a plain (not-highly-curved) dependence on $u$, the hasty shortcut may still exist if the relaxation modes change dramatically as one varies the control parameter $u$.

{\boldsymbol{$(\#3)$} }  states that $R_s(n)$ and/or $P_s$ occupies a large space in the $\vec p$-space. It increases the chance for the non-trivial intersection between $R_s$ and $P_s$, which allows for the existence of hasty shortcuts. Firstly, to allow $P_s$ to occupy a large space (i.e., greater dimensionality), it can be realized for systems whose eigenvectors consist of a large number of fast eigenvectors and only a few slow eigenvectors (i.e., $c$ is small in Eq.~\ref{eq:gap}). \footnote{One may conclude that a small $c$ value may result in a smaller reachable set $R_s$, which is partially dependent on the dimensionality of the slow manifold $\mathbb S(u)$'s. However, we show that one can increase the size of $R_s(n)$ by introducing a larger number of steps $n$.} Secondly, to increase the reachable set, one may consider increasing the number of steps in the control protocol. In other words, the larger the $n$, the larger the reachable set $R_s(n)$, as one can show $R_s(n)\subset R_s(n+1)$.

\begin{figure}[h]
    \includegraphics{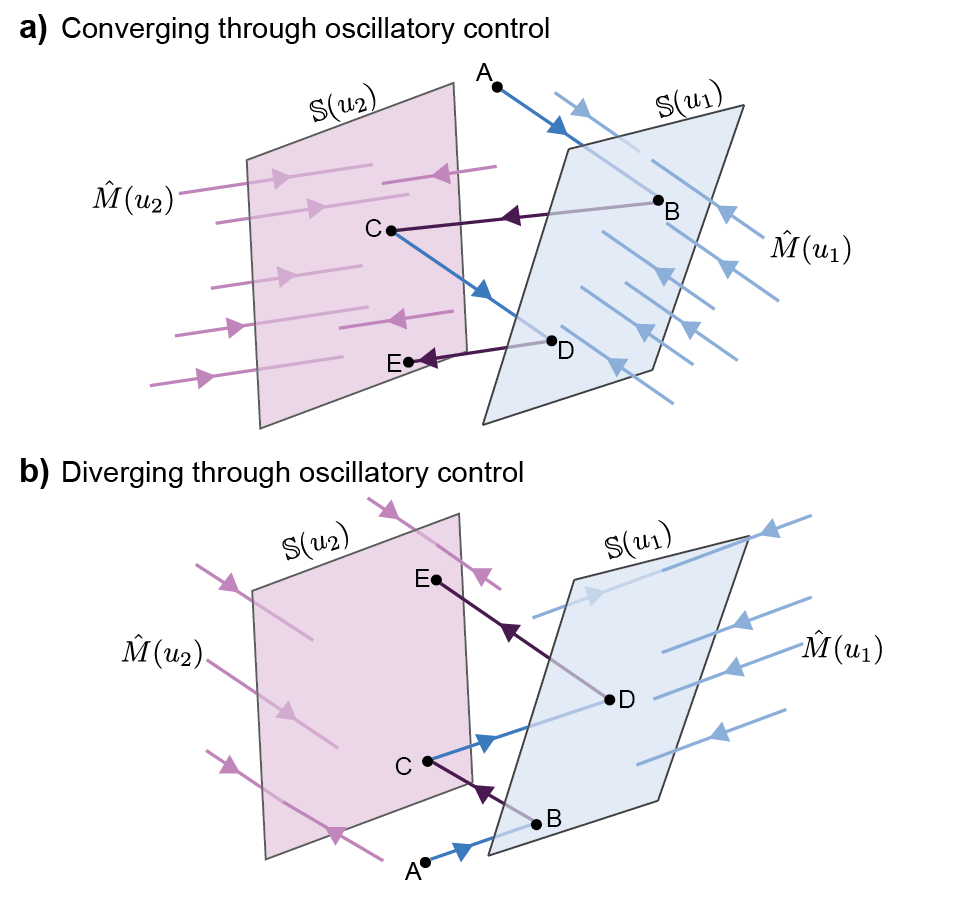}
    \caption{Illustration of effects of periodic control sequences. In each panel, the purple and blue planes represent two slow manifolds $\mathbb S(u)$ of two control values $u=u_1$ and $u=u_2$, and the solid arrows correspond to their corresponding fast projections $\hat M(u)$ for $u=u_1$ and $u=u_2$. In (a), an initial distribution A is driven by two oscillation periods ($u_1,u_2,u_1,u_2$) and evolves in a converging manner (see A,B,C,D,E). In (b), the initial distribution A is driven by two oscillation periods to evolve in a diverging manner (see A,B,C,D,E).}
    \label{fig:control}
\end{figure}

The increase of $R_s(n)$ as the number of steps $n$ increases can be analyzed geometrically. In Fig.~\ref{fig:control}, we illustrate two cases where increasing step number achieves new reachable states. For simplicity, consider a periodic control protocol oscillating between $u_1$ and $u_2$. Depending on the directions of the projection operators $\hat M(u_1)$ and $\hat M(u_2)$ relative to the locations of the slow manifolds $\mathbb S(u_1)$ and $\mathbb S(u_2)$, the system may be driven to a sequence of states that either converges or diverges. In Fig.~\ref{fig:control}a, as one increases the number of cycles between $u_1$ and $u_2$, the resulting states converge. In comparison, in Fig.~\ref{fig:control}b, the resulting system reaches a sequence of states that diverge as one increases the number of periods. In summary, geometric relations between the fast and slow eigenmodes for various values of $u$ could be engineered to increase the size of the reachable set $R_s(n)$ and thus increase the chance of obtaining hasty shortcuts.


\section{Numerical Simulation}
\label{sec:simulation}
\subsection{Assisted assembly model}
\label{subsec:assisted_assembly}
Let us now describe a simple Ising-type model of assisted assembly. Consider an octahedron-shaped lattice consisting of 8 empty sites, each of which can be occupied by up to one subunit. When a site $k$ is occupied, it is denoted by state $s_k=1$, and when empty, it is $s_k=0$. Thus, the configuration of each octahedron in the solution is represented by an $8$-bit string, and its energy is defined in an Ising model-like form. The energy of each configuration ${\bf s}=(s_1,\cdots,s_8)$ is given by the summation of the binding energy for each subunit and the interaction energy between neighboring subunits. 
\begin{equation}
\label{eq:ising}
E({\bf s}) = h \sum_k s_k + J\sum_{\langle l,m\rangle} s_l s_m
\end{equation}
where the constant $h$ is the binding energy between each subunit and the binding site, and $J$ is the interaction energy between any neighboring pairs of subunits.
If each binding site was distinguishable, the model consists of $2^8$ unique configurations. In reality, since each binding site is assumed to be identical with the same binding affinity with subunits, $h$, the $2^8$ configuration can be simplified to $24$ distinct configurations indexed by $i$, and their degeneracy is denoted by $g_i$. The degeneracy $g_i$ was calculated using six symmetry moves, and one can verify that $\sum_{i=1}^{24} g_i = 256$. Given this consideration, we can define the free energy of a distinguishable configuration $i$ as $F_i=E_i-\beta^{-1}\ln{g_i}$, where $\beta=1/ k_B T$ is assumed to be unity in this work.
\begin{figure}[h]
    \includegraphics{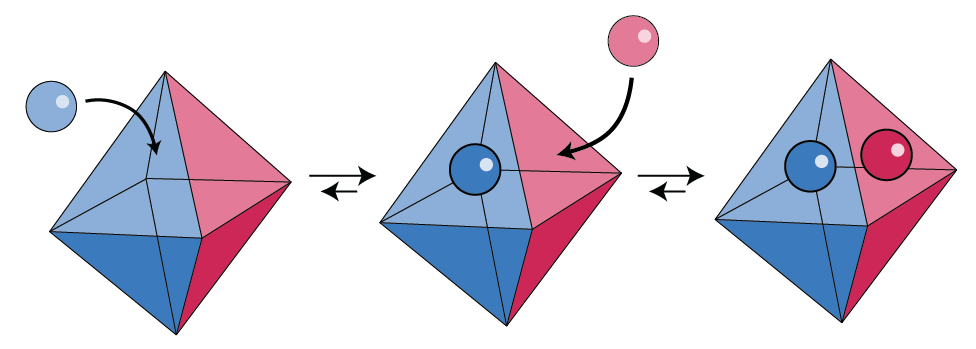}
    \caption{Simple assembly of an octahedral lattice model. Illustrated by two out of many possible transitions (binding and unbinding of a subunit).}
    \label{fig:lattice}
\end{figure}
We consider a dilute solution of octahedrons, where each octahedron is considered independent and identically distributed. Then the state of all octahedrons can be represented by the configuration probabilities of one octahedron. The probability $\vec p(t)$ evolves according to the concentration of subunits in the solution $c_s$, temperature $T$, and the interaction energies. Specifically, the probability evolves according to the master equation:
\begin{equation}
\frac{d \vec p(t)}{dt} = \hat R \cdot \vec p(t) 
\end{equation}
where $\hat R$ is the $24 \times 24$ transition matrix and its off-diagonal elements $R_{ij}$ represent probability transition rates from configuration $j$ to $i$. There are two types of transitions considered, binding or unbinding of a single subunit. The rate of any binding event is given as
\begin{equation}
\label{eq:Rcs}
R_{ji} = c_s e^{\beta\frac{F_j - F_i}{2}  + \beta \xi \eta(i,j)}
\end{equation}
and the rate of an unbinding event is given as 
\begin{equation}
\label{eq:Rnocs}
R_{ji} = e^{\beta \frac{F_j - F_i}{2} + \beta \xi \eta(i,j)}
\end{equation}
where $F_i$ denotes the free energy of configuration $i$. The term $\xi \eta(i,j)$ captures the steric hindrance barrier of the binding or unbinding transition due to occupied neighboring subunits. Here $\xi$ is a positive constant and $\eta(i,j)=\eta(j,i)$ is the occupation number of neighboring sites around the binding/unbinding site for the transition between configurations $i$ and $j$. In summary, the free energy barrier for the transition between configurations $i$ and $j$ is assumed to be their average free energy $(F_i+F_j)/2$ plus the steric hindrance $\xi \eta(i,j)$. One can verify that the transition rates satisfy the detailed balance condition. Some possible transitions are sketched in Fig.~\ref{fig:lattice}. The diagonal elements of the rate matrix are chosen such that each column of $\hat R$ adds up to 0. Thus the dynamics of assisted assembly can be described by a family of rate matrices $\hat R(c_s)$ parameterized by control parameter $c_s$. The theory of timescale separation and Eqs.~\ref{eq:master}, \ref{eq:solution}, \ref{eq:solution1} apply.

\subsection{Discrete reachable set and penultimate set}
\label{subsec:num_sets}
\begin{figure}[h]
    \includegraphics{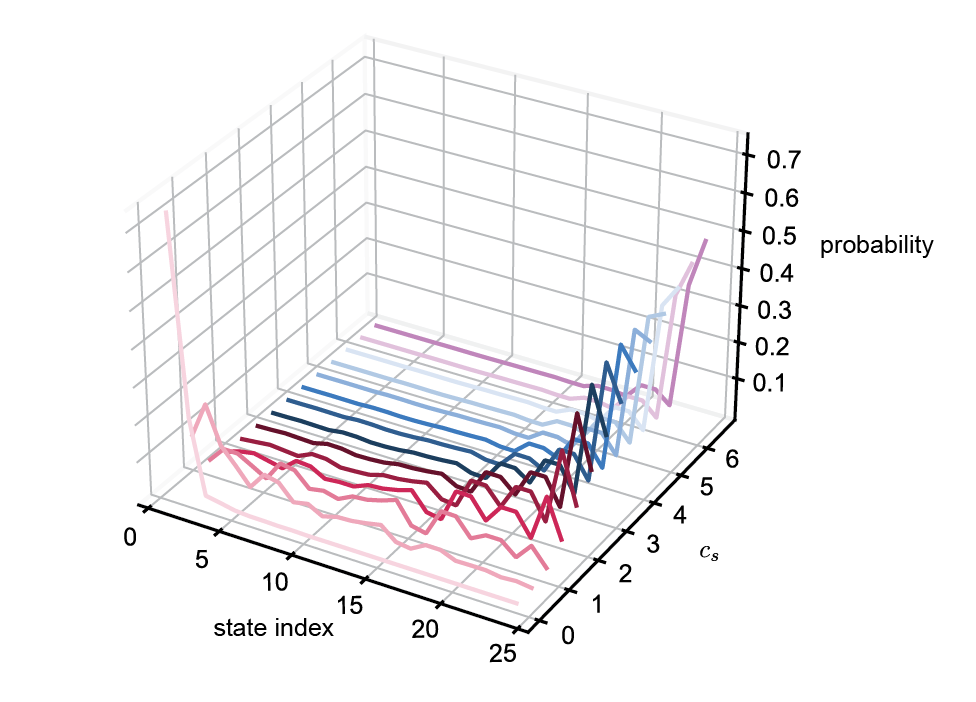}
    \caption{Equilibrium distributions for different assembly configurations at different values of subunit concentration $c_s$. There are $24$ distinct configurations of assembly denoted by state index. Larger state indexes correspond to higher-occupancy assembled configurations.}
    \label{fig:equilibra}
\end{figure}

The system's dynamics at a given subunit concentration $c_s$ is captured by the corresponding rate matrix $\hat R(c_s)$. The control parameter $u$ for this assembly problem is chosen to be the subunit concentration $c_s$. To illustrate the systems' equilibrium distributions at different control values, we discretized $c_s$ and obtained the equilibrium distribution for each $c_s$, shown in Fig.~\ref{fig:equilibra}.

Numerical eigenanalysis of $R(c_s)$ reveals that the system shows timescale separation over an arbitrary selection of values of $c_s$ (see Fig.~\ref{fig:eval_crossing}). Therefore, we can use the eigenmode separation approach described in Sec.~\ref{subsec:time_sep} to perform an efficient simulation of the system's dynamics in a reduced space. It demonstrates good agreement between the full dynamics and the reduced dynamics based on time-scale separation (see Fig.~\ref{fig:full_vs_reduced}). 

\begin{figure}[h]
    \includegraphics{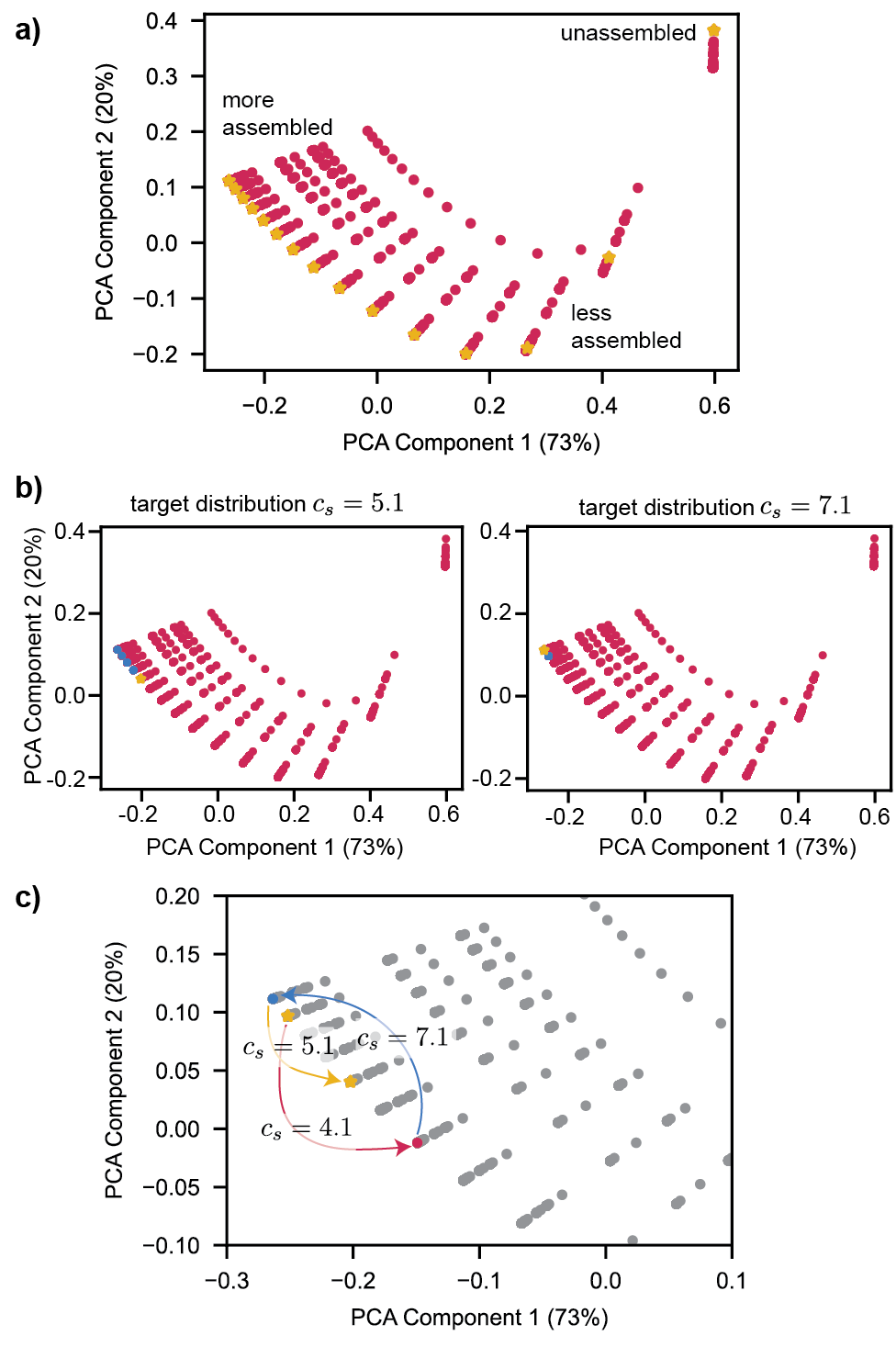}
    \caption{Numerical generation of reachable and penultimate sets. a) The reachable set (red dots) in a 24-dimensional state space is here represented in reduced dimensions using PCA. Relevant equilibrium distributions (golden stars) correspond to equilibria previously shown in Fig.~\ref{fig:equilibra}. b) Penultimate sets (blue dots) corresponding to two different equilibria characterized by $c_s$ (golden stars) are shown and indicate differences in reachability depending on the choice of final distribution. c) A representative trajectory of nonequilibrium driving between two equilibria. Fast dynamics mapping (arrows) connect the initial equilibrium (higher golden star) first to a member of the reachable set (red dot), then to a member of the penultimate set (blue dot), and arrive at the final equilibrium (lower golden star) using its associated fast projection.}
    \label{fig:set_generation}
\end{figure}
With a numerically solvable model, we verify our geometrical theory for the existence of hasty shortcuts. We simplify the control protocol by requiring the values of $c_s$ to be chosen from a discrete set $c_s=0.1,~0.6,~\cdots,~7.1$. In the rest of the paper, $u$ and $c_s$ both denote the control parameter and will be used interchangeably. Then by further discretizing the probability distribution space, we numerically obtain the discrete reachable set (as dots in Fig.~\ref{fig:set_generation}a) and its nontrivial intersection with the penultimate set (as blue dots in Fig.~\ref{fig:set_generation}b). 
\begin{figure*}
    \includegraphics{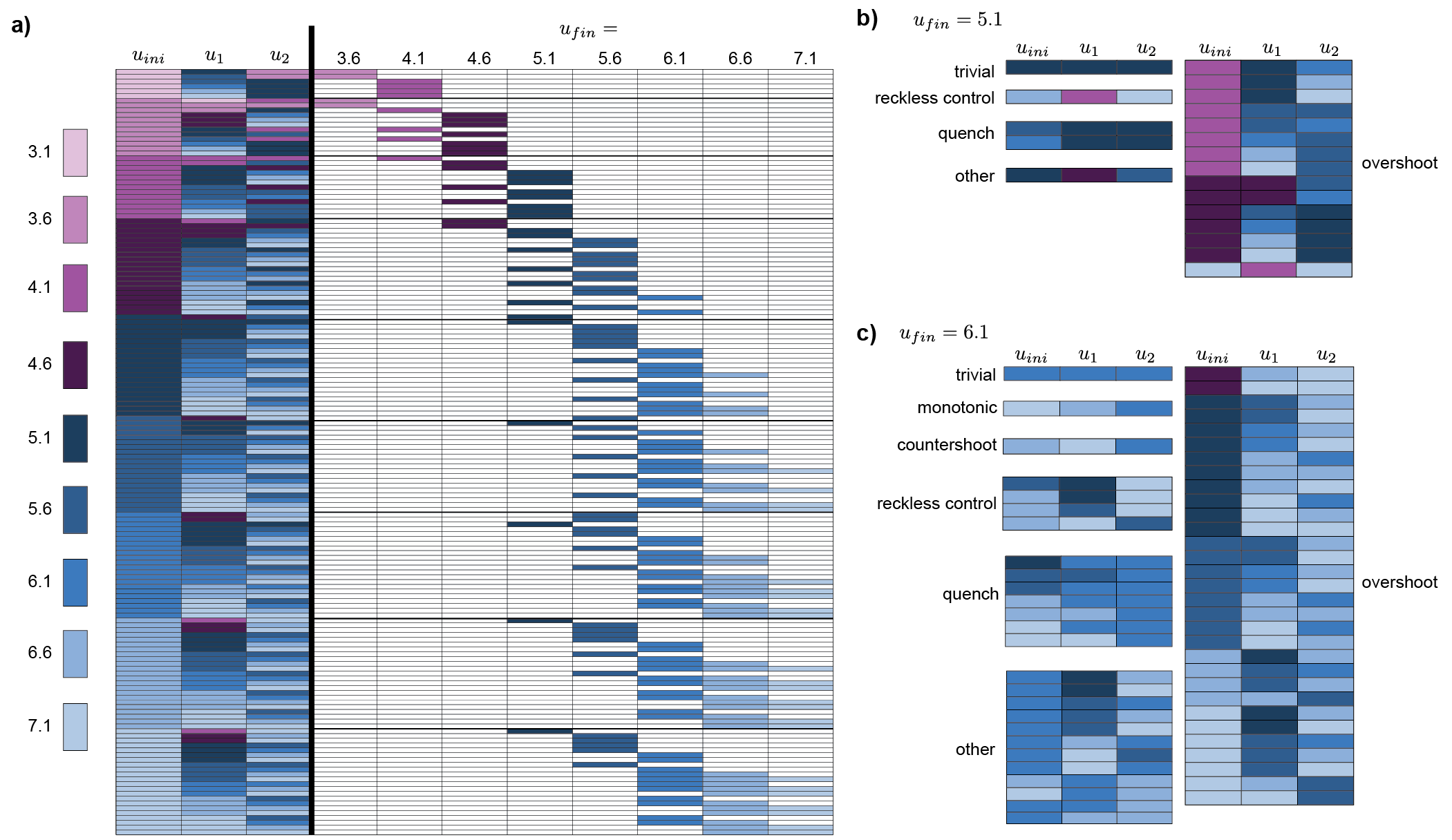}
    \caption{Complete list of $(n=2)$-step hasty shortcuts. a) The control protocols $(u_{ini},u_1,u_2,u_{fin})$ are shown by the combination of the left three columns ($u_{ini},u_1,u_2$) and the eight columns on the right ($u_{fin}$). If a final steady state for a given $u_{fin}$ is reachable following any control shown in the left three columns ($u_{ini},u_1,u_2$), then it is illustrated by a shaded box for the corresponding ultimate control $u_{fin}=3.1,~3.6,~\cdots,~7.1$. b) The control protocols that can reach $u_{fin} = 5.1$ are represented by the corresponding ($u_{ini},u_1,u_2$), and they are grouped by trivial (stationary control, not a shortcut), quench, reckless, overshoot, and other categories. c)The control protocols that can reach $u_{fin} = 5.1$ are represented by the corresponding ($u_{ini},u_1,u_2$), and they are grouped by trivial, monotonic, countershoot, reckless control, quench, overshoot, and other categories.}
    \label{fig:patterning}
\end{figure*}
The reachable set is constructed as follows. Consider the system starts at a steady-state probability distribution $\vec p(t_0)=\vec p^{~ss}_{u_{ini}}$ for the initial control parameter to be the subunit concentration chosen from the discrete list $c_s=0.1,~0.6,~\cdots,~7.1$. We perform two rapid steps of controls $(u_1, u_2)$ to drive the system. After the two rapid projections, the system reaches $\vec p(t_2)= \hat M(u_1) \hat M (u_2) \vec p^{~ss}_{u_{ini}}$. By choosing all possible protocols $(u_{ini},u_1, u_2)$, we can steer the system's probability distributions into a two-step reachable set $R_s(2)$. Notice that each element of the set is a $24$-dim vector in the probability simplex, and for illustrative purposes, they are shown in a $2$-dim plot by principal component analysis (PCA) in Fig.~\ref{fig:set_generation}a. Notice that if $u_{ini}=u_1=u_2$, the control protocol does not change the parameter at all, and the system remains at the initial equilibrium distribution. The reachable set contains the $15$ steady-state distributions corresponding to each discrete value of $c_s$, shown by $15$ golden stars in Fig.~\ref{fig:set_generation}a.

To determine the hasty shortcuts, we numerically find the intersection between the reachable set and the penultimate set. This can be done by taking any probability distribution from the reachable set, e.g., $\vec p(t_2)$, applying an ultimate-step control of $u_{fin}$. If the rapid relaxation $\hat M(u_{fin})$ projects $\vec p(t_2)$ to the steady state $\vec p^{~ss}_{u_{fin}}$, then the distribution $\vec p(t_2)$ (from the reachable set) also belongs to the penultimate set. Numerically, the intersection between a point $\vec p(t_2)$ in the reachable set and a point in the penultimate set (corresponding to $u_{fin}$) is judged by the following: if $max \left (\hat M(u_{fin}) \vec p(t_2) - \vec p^{~ss}_{u_{fin}}\right )<\epsilon = 10^{-7}$, then $\vec p(t_2)$ can be mapped by a final control $u_{fin}$ to a state that is extremely close to the ultimate steady state, $\vec p^{~ss}_{u_{fin}}$. Moreover, the corresponding control sequence, $(u_{ini},u_1, u_2,u_{fin})$, constructs a hasty shortcut connecting the initial steady state $\vec p^{~ss}_{u_{ini}}$ to the final steady state $\vec p^{~ss}_{u_{fin}}$, which only utilizes the fast dynamics and does not require waiting for any slow dynamics. Notice that rather than showing the complete overlap between the reachable set and the penultimate set, in Fig.~\ref{fig:set_generation}b and Fig.~\ref{fig:penultimate_set}, we have shown the intersection set corresponding to different choices of final control values $u_{fin}$. In Fig.~\ref{fig:set_generation}b left, the intersection set contains four different distributions for $u_{fin}\rightarrow c_s=5.1$. It indicates that the steady-state distribution corresponding to $c_s=5.1$ can be accessed through different members of the penultimate set. In contrast, for the final steady state $\vec p_{ss}(c_s = 7.1)$, only a single non-trivial intersection exists (see blue dots shown in Fig.~\ref{fig:set_generation}b right).

As demonstrated above, even in a simple self-assembly system, we can identify multi-step and non-monotonic hasty shortcuts capable of driving an initial equilibrium distribution to a desired final equilibrium. One such shortcut is illustrated in Fig.~\ref{fig:set_generation}c where a system is driven from $\vec p_{ss}(c_s = 6.6)$ to $\vec p_{ss}(c_s = 5.1)$. This non-monotonic control involves starting at the steady state of concentration $u_{ini}=6.6$, applying a sudden decrease of subunit concentration to $u_1 = 4.1$, immediately followed by a dramatic subunit concentration increase, $u_2 = 7.1$, and ultimately switching the subunit concentration to the desired final value $c_s=5.1$. This shortcut is one of many hasty shortcuts obtained by the reachable set-penultimate set intersection.

\subsection{Classification of nonequilibrium hasty shortcuts}
\label{subsec:num_hasty}
Even for a low number of control steps $n=2$, the system demonstrates numerous hasty shortcuts. In Fig.~\ref{fig:patterning}a, we list all $(n=2)$-step hasty shortcuts starting from initial steady state of $u_{ini}= 3.1,~ 3.6,~ \cdots, 7.1$ and  ending at a final steady state of $u_{fin}$, i.e., $u_{ini}, u_1, u_2,u_{fin}$. For illustrative purposes, the shortcuts are represented by two parts: (1) the initial three control values $u_{ini},u_1,u_2$ are shown as the three columns to the left of the vertical bold line, and (2) then each ultimate control $u_{fin}$ that can lead the system, prepared by an initial control to reach $\vec p^{~ss}_{u_{fin}}$ is shown as a filled (colored) block in the eight columns to the right side of the vertical bold line. The color of each block corresponds to the value of $u$. Notice that among all possible rapid control sequences of $n=2$, many cannot achieve any ultimate equilibrium and thus do not qualify as shortcuts. The complete list of all possible control sequences is presented in Fig.~\ref{fig:all_two_step_trajectories}. Also, we notice that there does not exist any $(n=2)$-step hasty shortcut starting from or ending at low initial concentrations ($u_{ini}<3.1$ or $u_{fin}<3.1$), which indicates that a large number of steps $n$ may be required to find shortcuts in low concentration region.  

The numerical results revealed numerous hasty shortcuts of diverse temporal patterns. Here we classify the shortcut into a few categories of interest. Notice that the Mpemba-effect-like shortcuts only constitute a small fraction of all possible hasty shortcuts. We classify hasty shortcuts as (i) monotonic; (ii) overshooting only; (iii) counter-shooting only; (iv) reckless control; (v) quenching, and (vi) other:
\begin{itemize}
    \item $(i)$ ``monotonic'', the shortcut involves $u_1$ and $u_2$ that are within the closed range of $[u_{ini},u_{fin}]$ (or $[u_{fin},u_{ini}]$) and the control is monotonic, $u_1\leq u_2$ (or $u_1\geq u_2$).
    \item $(ii)$ ``overshooting only'', one or more intermediate control $u_1$ or $u_2$ go beyond the range between $u_{ini}$ and $u_{fin}$ along the forward direction (from $u_{ini}$ to $u_{fin}$), and thus creates an overshoot.
    \item $(iii)$ ``counter-shooting only'', one or both intermediate controls $u_1$ of $u_2$ go beyond the range between $u_{ini}$ and $u_{fin}$ along the counter-forward direction (from $u_{fin}$ to $u_{ini}$). By definition, the Mpemba-effect-like shortcuts must belong to this category.
    \item $(iv)$ ``reckless control'', one of the intermediate control (either $u_1$ or $u_2$) overshoots, and the other intermediate control counter-shoots. This is reckless as it involves large-amplitude controls beyond $u_{ini}$ and $u_{fin}$ in both the forward and backward directions. 
    \item $(v)$ ``quenching'', the control is a direct 1-step quench from $u_{ini}$ to $u_{fin}$ that can immediately drive the system from initial equilibrium to the final equilibrium. 
    \item $(vi)$ ``other'' involves all kinds of controls that do not fall in the classifications above. 
\end{itemize}

In Fig.~\ref{fig:patterning}b-c, we illustrate the classification of shortcuts by presenting the hasty shortcuts that can reach two specific final equilibria (i.e., for $u_{fin}=5.1$ and $u_{fin}=6.1$). It is interesting to note that no monotonic or counter-shoot shortcuts can arrive at the steady state of $u_{fin}=5.1$. Still, all categories of shortcuts are found to drive some initial equilibrium to the final equilibrium of $u_{fin}=6.1$. Also, for certain pairs of initial $u_{ini}$ and final $u_{fin}$, there may be no direct quenching or monotonic shortcut. In Fig.~\ref{fig:set_generation}c, only two shortcuts are found to start from $u_{ini}=4.6$ and end at $u_{fin}=6.1$, and both are overshooting shortcuts. For a complete summary of all shortcuts for all possible final equilibria, please refer to Fig.~\ref{fig:two_step_classes}.

\begin{figure}[h]
    \includegraphics{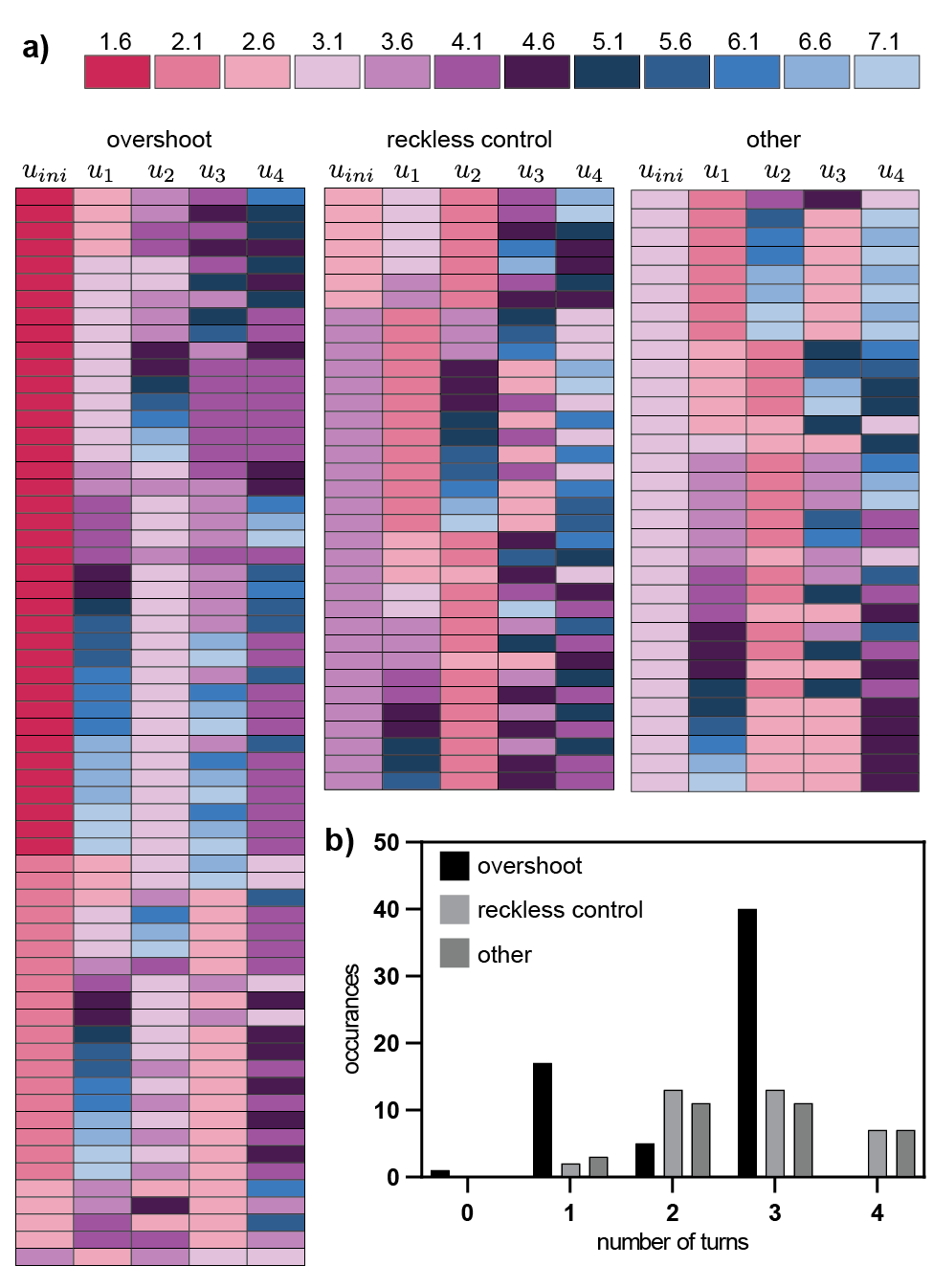}
    \caption{All $(n=4)$-step hasty shortcuts that can reach final equilinrium of $u_{fin} = 3.1$. a) Shortcuts are grouped into overshoot, reckless control, and other categories. b) Histograms of direction turns for hasty shortcuts in each category.}
    \label{fig:multistep}
\end{figure}

We further demonstrate that by increasing the number of steps $n$, one can reveal more hasty shortcuts by achieving a larger reachable set $R_s(n)$ as argued in the geometric feature $(\# 3)$ in Sec.~\ref{subsec:geo_eigen}. The low concentration region $c_s<3.1$, which is inaccessible by $(n=2)$-step shortcuts (see Fig.~\ref{fig:patterning}), can be accessed by $(n=4)$-step shortcuts (see Fig.~\ref{fig:multistep}). To demonstrate, we numerically construct the reachable set $R_s(4)$ by using $\hat M(u_4) \hat M(u_3)\hat M(u_2)\hat M(u_1) p^{~ss}_{u_{ini}}$, where the initial equilibria are restricted to the low concentration region $u_{ini} =c_s = 0.1, \cdots, 3.6$. A list of new hasty shortcuts is found to connect from $u_{ini}\leq 3.6$ to any ultimate value $u_{fin}$ is shown in Figs.~\ref{fig:four_step_trajectories}-\ref{fig:four_step_classes}. In Fig.~\ref{fig:multistep}a, we list all hasty shortcuts toward $\vec p^{~ss}_{u_{fin}}$ for $u_{fin} = 3.1$. 

In this low concentration range, where $(n=2)$-step shortcuts do not exist, $(n=4)$-step shortcuts are also harder to achieve: As shown in Fig.~\ref{fig:multistep}a, control protocols with simple patterns, such as quench, undershoot, and monotonic, cannot serve as hasty shortcuts. Only three categories of shortcuts are observed: overshoot, reckless control, and others.  
This observation agrees with the geometric analysis following feature $(\#3)$ in Sec.~\ref{subsec:geo_eigen}, where erratic or oscillatory multi-step control protocols with a number of direction turns can lead to shortcuts where simpler patterns fail to reach. Here a direction turn is defined by one change of control direction of $u_{i-1},u_i,u_{i+1}$, e.g., $u_1 > u_2 < u_3$ or $u_1 < u_2 > u_3$. By quantifying the number of direction turns in each $(n=4)$-step hasty shortcut, we find that many overshoot-type shortcuts are strongly ``oscillatory'', involving three direction turns. We also observe a number of reckless control or other shortcuts that involve two, three, or four direction turns. The histograms of the number of direction turns in each category of shortcuts are shown in Fig.~\ref{fig:multistep}b.

\begin{figure}[h]
    \includegraphics{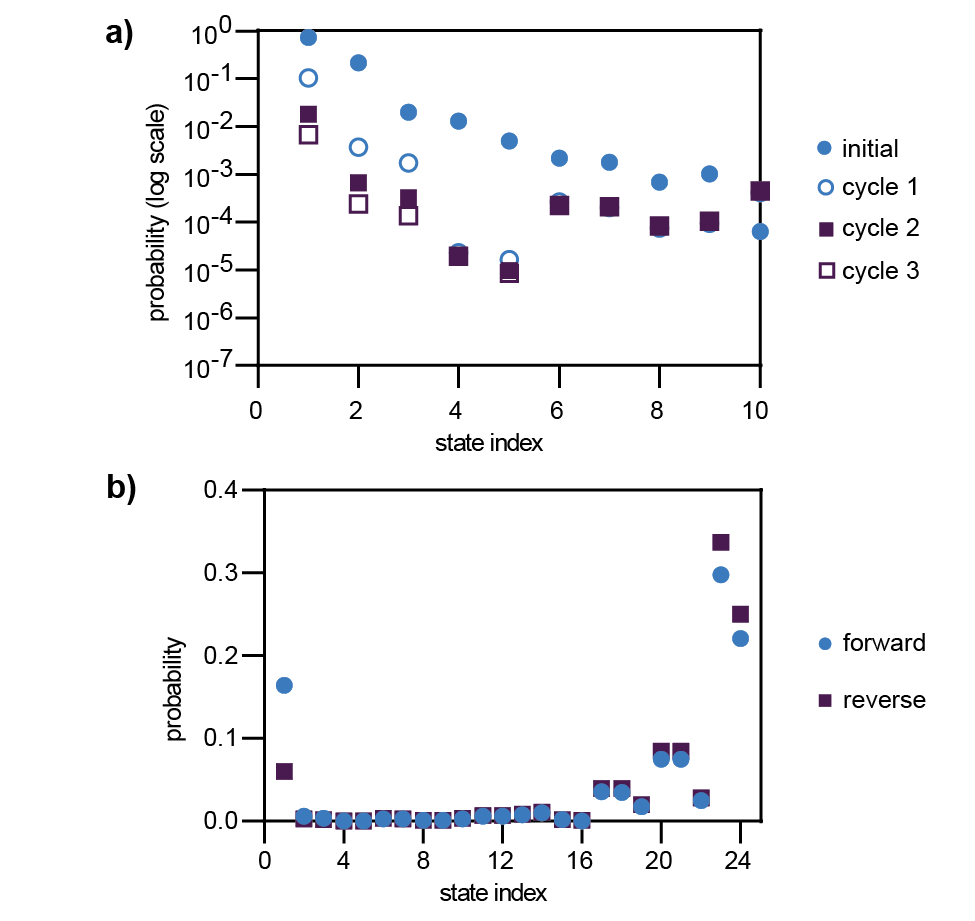}
    \caption{Effect of periodic controls and the noncommutative nature of the evolution a) Repeated periodic oscillations between two different $u$ shows probability distributions evolving convergingly. b) For a cycle of control $u_1 = 0.1$, $u_2 = 7.1$, and $u_i = 3.6$, the forward cycle and the reverse cycle result in different ultimate distributions. 
    }
    \label{fig:cycles}
\end{figure}

We further investigate the effect of control protocol with strictly periodic oscillations between $u_1$ and $u_2$ (i.e., periodic direction turns). We choose an initial equilibrium of $u_{ini} = 0.1$ and then oscillate the control parameter between $u_1 = 7.1$ and $u_2 = 1.1$. In Fig.~\ref{fig:cycles}a, we demonstrate the evolution of the initial distribution after a single period($\hat M(u_1)\hat M(u_2)$), double periods ($[\hat M(u_1)\hat M(u_2)]^2$), and triple periods ($[\hat M(u_1)\hat M(u_2)]^3$). We verify that the system evolves into a different distribution at the end of each period in a converging manner as described by Fig.~\ref{fig:control}a in Sec.~\ref{subsec:geo_exist}. In Fig.~\ref{fig:cycles}b, we also demonstrate the noncommutative property of $(u_1, u_2)$ pairs in Fig.~\ref{fig:cycles}b, where $\hat M(u_i) \hat M(u_2) \hat M(u_1) \vec p^{~ss}_{u_i}$ (forward cycle) and $\hat M(u_i) \hat M(u_1) \hat M(u_2) \vec p^{~ss}_{u_i}$ (backward cycle) lead to different ultimate states. 
The periodic oscillatory and noncommutative control analysis further underscores the importance of oscillatory, nonmonotonic, and reckless patterns in achieving hasty shortcuts.

\section{Conclusions}
\label{sec:conclusion}
This paper proposes nonequilibrium hasty shortcuts, a new class of rapidly steered control protocols that can manipulate a system from an initial equilibrium (or steady state) to a different final equilibrium (or steady state). The name ``hasty'' indicates that throughout the process, the system's dynamics only involve fast dynamics but not slow relaxations (due to the rapid change of the control parameter). We have described a general geometric approach to understanding the nonequilibrium hasty shortcuts in arbitrary systems that can be described by a master equation. At any transiently fixed control parameter $u$, the fast dynamics ($\hat M(u)$) rapidly projects any initial probability distribution onto a low dimensional slow manifold $\mathbb S(u)$ that is spanned by the slow relaxation modes. By allowing the control parameter $u$ to assume a range of values, we obtain a family of slow manifolds and a family of fast projection operators. This allows us to construct a reachable set and a penultimate set in the probability space. Moreover, the existence of hasty shortcuts can be determined through the nontrivial intersection between the reachable set and the penultimate set. The geometric perspective of this paper allows us to illustrate the possible mechanisms that could allow a system to have hasty shortcuts.  

As an illustration, we demonstrate hasty shortcuts in a simple Ising-model-like assisted assembly process where each assembly is an 8-sided lattice. Each lattice site can bind up to one subunit. The control parameter $u$ is the concentration of the subunit in the solution $c_s$. We numerically generate the reachable and penultimate sets and demonstrate many nonequilibrium hasty shortcuts of various patterns and step lengths. Tabulating the complete set of shortcut protocols revealed a diverse set of shortcuts, including monotonic, overshooting, countershooting, reckless control, quenching, and many others. Among these shortcuts, the countershooting shortcuts may resemble the strong Mpemba effect that was previously predicted \cite{klich2019mpemba, Gal2020-yq}. 

This geometric approach lays the foundation for future research into the processes that make hasty shortcuts possible, thus enriching the field of nonequilibrium thermodynamic control strategies. In the future, such a geometric approach may be extended to controlling non-adiabatic quantum dynamics and strongly dissipative systems.

\section*{Supplementary Information}
Further results on the numerical simulation can be found in the Supplementary Information.

\begin{acknowledgements}
The authors appreciate the inspiring discussions with Dr. Zhongmin Zhang and Prof. Christopher Jarzynski. S.S.C. acknowledges the National Science Foundation Graduate Research Fellowship (DGE-2040435). We also appreciate the financial support from the startup
fund at UNC-Chapel Hill and the fund from the National Science Foundation Grant No. DMR-2145256.
\end{acknowledgements}

\appendix

\section{Simulation Details}
\subsection{Master equation dynamics}
The assembly system used in this study is a solution of tetrahedrons. Each tetrahedron serves as an 8-sited lattice. Each site can be bound with up to one subunit from the solution. The concentration of subunits is considered a controlled parameter (by a large bath in contact with the solution) and is denoted by $c_s$. Each site $k$ has two possible states: $s_k=1$ when occupied by a subunit and $s_k=0$ when unoccupied. As a result, the configuration of each tetrahedron is described by a string of bits $\textbf{s} = (s_1, \cdots, s_8)$. The energy of each configuration is given by the following Ising-type form, according to Eq.~\ref{eq:ising} in the main text:
\begin{equation}
E({\bf s}) = h \sum_k s_k + J\sum_{\langle l,m\rangle} s_l s_m
\end{equation}
where $h=1$, and $J=-0.5$ and the units are chosen such that $k_BT=1$.
Assuming that the tetrahedrons form a dilute solution, where we can describe the self-assembly dynamics by the probability distribution of a single tetrahedron over its configuration space. Here we have taken advantage of the fact that in the dilute solution, each tetrahedron are independent of each other yet follows an identical probability distribution. The probability distribution evolves according to the Master Equation:
\begin{equation}
    \frac{d \vec p}{dt} = \hat R \cdot \vec p(t)
\end{equation}
where $\vec p(t)$ is the probability vector at time $t$ among all configurations and $\hat R$ is the transition rate matrix. Each element of $\hat R$, $R_{ij}$, describes the transition rate of configuration $j$ transitioning to configuration $i$, which corresponds to an event of single-subunit binding or unbinding from one lattice site. The transitions rate, illustrated for an unbinding transition, $R_{ij}$ can be obtained by the Arrhenius rate:
\begin{align}
    R_{ji} &= e^{\beta (B_{ji} - F_i)} \\
    &= e^{\beta\left (\frac{F_j + F_i}{2} + \xi \eta - F_i \right )} \\
    &= e^{\beta \left (\frac{F_j - F_i}{2} + \xi \eta\right )} 
\end{align}
where the barrier height is $B_{ij} = \frac{F_j + F_i}{2}+ \xi\eta(i,j)$, and $\eta(i,j)$ is the number of occupied neighboring sites to the binding or unbinding site. Notice that the positive constant $\xi$ indicates that it is more difficult to bind or unbind at a site if there are many neighboring sites in the bound state. The $\xi\eta$ term in the barrier resembles the steric hindrance discussed in organic chemistry. In our simulations, we take $\xi = 4$ and use the unit convention where $\beta = 1$. If this transition is a binding event, its rate is given according to Eq.~\ref{eq:Rcs} in the main text:
\begin{equation}
\label{eq:A6}
R_{ji} = c_s e^{\beta\frac{F_j - F_i}{2}  + \beta \xi \eta(i,j)}
\end{equation}
If this transition is an unbinding event, its rate is given according to Eq.~\ref{eq:Rnocs} in the main text:
\begin{equation}
\label{eq:A7}
R_{ji} = e^{\frac{F_j - F_i}{2} + \beta \xi \eta(i,j)}
\end{equation}
In either case, this rate depends on the free energy $F_i$ and $F_j$ of the configurations, and $\xi \eta(i,j)$ is the steric hindrance term. We determine  The free energy of any configuration $i$ is given as
\begin{equation}
    F_i = E_i - T \ln g_i
\end{equation}
where $g_i$ is the degeneracy of configuration $i$. This degeneracy was determined using 6 non-unique symmetry moves involving 5 types of reflections and one type of rotation. We therefore determined the set of geometrically unique configurations, $\mathbb C$, as follows. The state indexed by ``1'' refers to the unique all-empty configuration with zero subunits bound to the tetrahedron, which has degeneracy $g=1$. The state indexed by ``2'' refers to a configuration with one subunit bound, which has degeneracy $g = 8$. Degeneracies of all other states were determined using the following algorithm:
\begin{algorithm}[H]
    \begin{algorithmic}[1]
        \State Initialize an unoccupied configuration
        \State Occupy 1 position with degeneracy $g=8$
        \State Pass to configuration set $\mathbb C$
        \For{$i=2:8$}
            \For{$s$ in $\mathbb C$}
                \State Find all unoccupied positions $s_k$ in $s$
                \For{$(s_k)_i$ in $s_k$}
                    \State Occupy position $(s_k)_i$
                    \State Generate all configurations $s_{k,test}$ by symmetry rules
                    \State Remove duplicates in $s_{k,test}$
                        \For{$s_{k,i}$ in $s_{k,test}$}
                            \State Search for any matches in $\mathbb C$
                                \If{match found}
                                    \State Calculate transition rate
                                    \State Assign matrix element
                                \Else
                                    \State Pass new configuration to $\mathbb C$
                                    \State Size of $s_{k,test}$ is degeneracy $g$
                                    \State Calculate transition rate
                                    \State Assign matrix element
                                \EndIf
                        \EndFor
                \EndFor
            \EndFor
        \EndFor
    \end{algorithmic}
\end{algorithm}

Using this algorithm, we identified $N = 24$ geometrically distinct configurations and further verified the total number of degeneracies in the system $2^8 = 256$. 

Once all possible $R_{ij}$ were obtained by Eqs.~\ref{eq:A6} and \ref{eq:A7}, the diagonal elements $R_{ii}$ were calculated using 
\begin{equation}
    R_{ii} = - \sum_k R_{ki}
\end{equation}
which ensures that the master equation preserves normalization. One can show that the leading eigenvalue of $\hat R$ is 0, with the real part of all other eigenvalues strictly $< 0$. 

\subsection{Construction of fast dynamics projection operator $\hat M$}
\begin{figure}[H]
    \centering
    \includegraphics[scale=1]{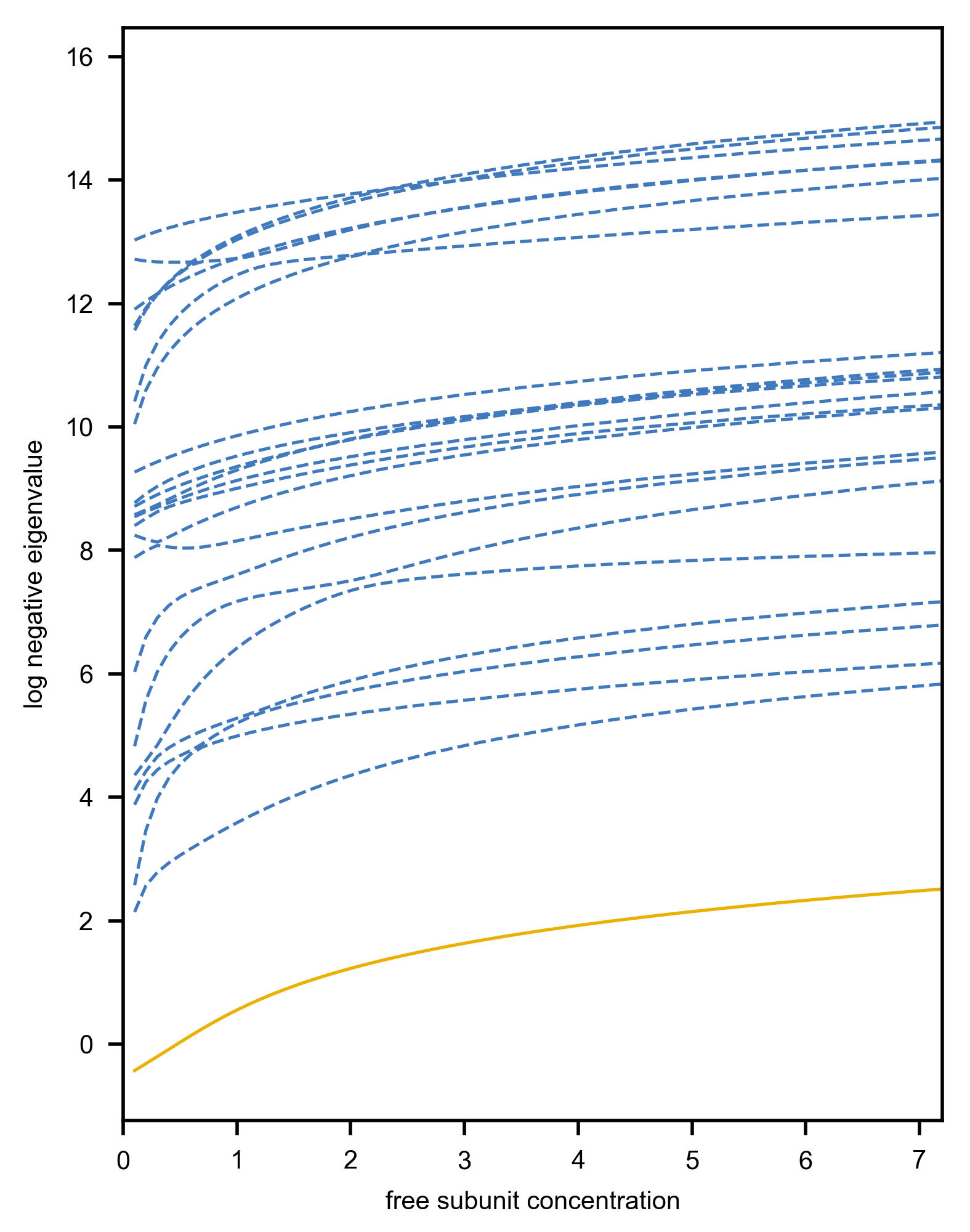}
    \caption{Eigenvalues of $\hat R(c_s)$ for a range of $c_s$. The yellow solid line represents the leading nonzero eigenvalue corresponding to the slowest relaxation mode. The dashed blue lines correspond to faster relaxation modes. The zero eigenvalue is not shown here.}
    \label{fig:eval_crossing}
\end{figure}

The matrix $\hat R$, which depends on the free subunit concentration $c_s$, can be decomposed into
\begin{equation}
    \hat R(c_s) = \hat V \hat \Lambda [\hat V]^-1
\end{equation}
where $\hat V$ is a matrix containing right eigenvectors $\vec v_i, \cdots, \vec v_n$ on the columns and $\Lambda$ is a diagonal matrix containing the eigenvalues $\lambda$. Closer examination revealed that there is timescale separation in the eigenvalue spectrum for the whole range of $c_s$ (see Fig.~\ref{fig:eval_crossing}).

If neighboring $\lambda_i$ and $\lambda_{i+1}$ exhibit at least a 10-fold difference, the cutoff lambda $\lambda_c$ is designated to be $\lambda_i$. In most cases, at any chosen $c_s$, the cutoff was $c = 2$. The eigenmode corresponding to $\lambda_c$, $\vec v_c$, contains information about the support along which slow mode dynamics evolve. The support for the slow modes manifold corresponded to states in $\vec v_c$ that had a magnitude greater than 0.05.

The fast dynamics operator $\hat M$, given according to Eq.(8) in the main text, projects the full $N$-state space onto the support of the slow dynamics. This matrix was constructed by first writing down the generator
\begin{equation}
    \hat W_{\tau} = e^{\hat R \tau}
\end{equation}
where $\hat R$ is the full transition rate matrix and $\tau = \frac{-1}{\lambda_c}$. This projection operator encompasses fast dynamics within the system. If one is looking to propagate this system in a reduced state space, this mapping matrix can be constructed and normalized as a non-square matrix with number of rows equal to $N$ and number of columns equal to the size of the support. The full $N$ state space can be recovered after the projection by inserting equilibrium values of all states not contained in the support.

\begin{figure}[h]
    \centering
    \includegraphics{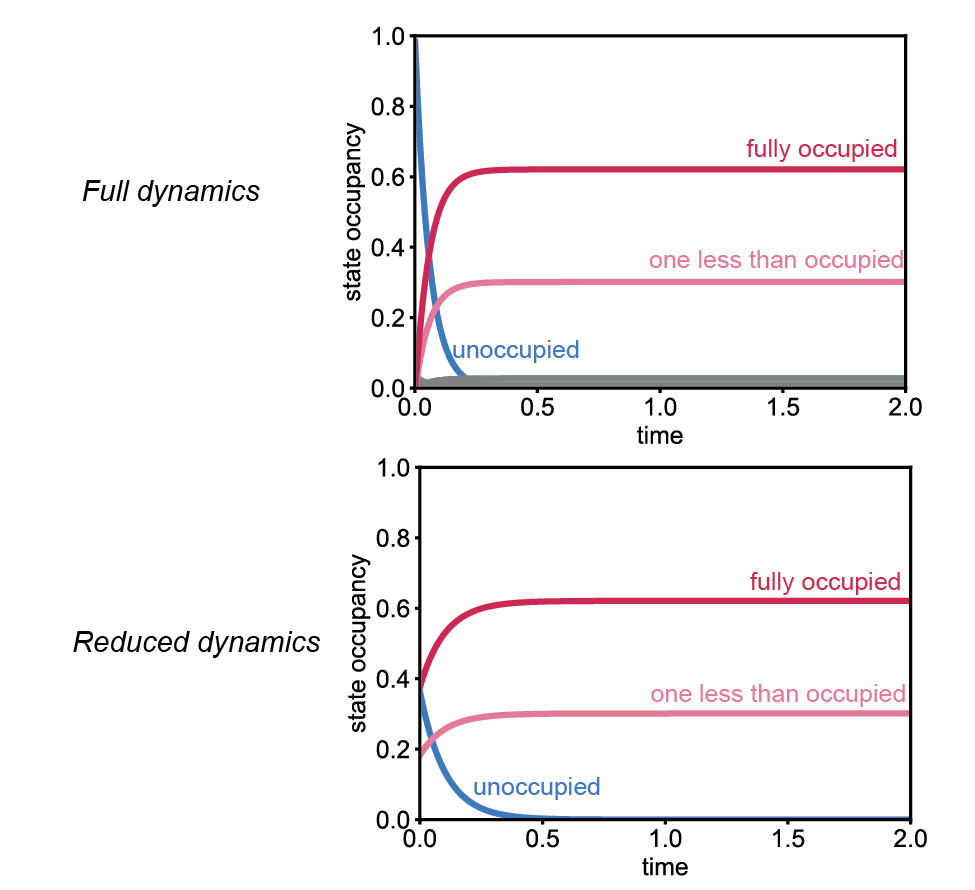}
    \caption{Comparison of full dynamics (top pane) and reduced dynamics (bottom pane) at $c_s = 10$. States contained in the slow manifold are shown in both plots, including unoccupied state (blue) and two occupied states (red and light red). The full dynamics also handles states not included in the support (gray).}
    \label{fig:full_vs_reduced}
\end{figure}

The slow dynamics operator evolves a state only along the support using dynamics of eigenvalues greater than $\lambda_c$. This matrix was constructed by writing down the slow dynamics generator:
\begin{equation}
    \hat W_{\tau} = (e^{\hat R * \delta t})^{1000}
\end{equation}
where $\delta t = \frac{-1}{1000 \lambda_c}$. Using this generator, a slow dynamics operator was constructed and normalized that acts on states contained within the slow modes support. We validated the construction of both of these operators by comparing the reduced dynamics with the full dynamics and demonstrating good agreement (Fig.~\ref{fig:full_vs_reduced}).

\section{Numerical shortcut generation and results}
\subsection{Reachable and penultimate sets}
We constructed the reachable set $R_s$ described by Eq.~\ref{eq:Rs} in the main text by first designating the concentration of free subunit $c_s$ to be the control parameter $u$. We discretized the $c_s$ into 15 distinct environments to form the set $\{ c_s \}$. We constructed $\hat R(c_s)$ for each and evaluated the equilibrium distribution $\vec p^{~ss}_{c_s}$. We then evaluated
\begin{equation}
    \hat M(u_1) \hat M(u_2) \vec p^{~ss}_{u_{ini}}
\end{equation}
for each possible environment pattern. This approach was used to construct $R_s$. We represent $R$ in a reduced 2-dimensional state space using Principle Component Analysis (PCA, sklearn package from Python 3.9).

\subsection{Temporal pattern analysis}
\begin{figure*}
    \centering
    \includegraphics{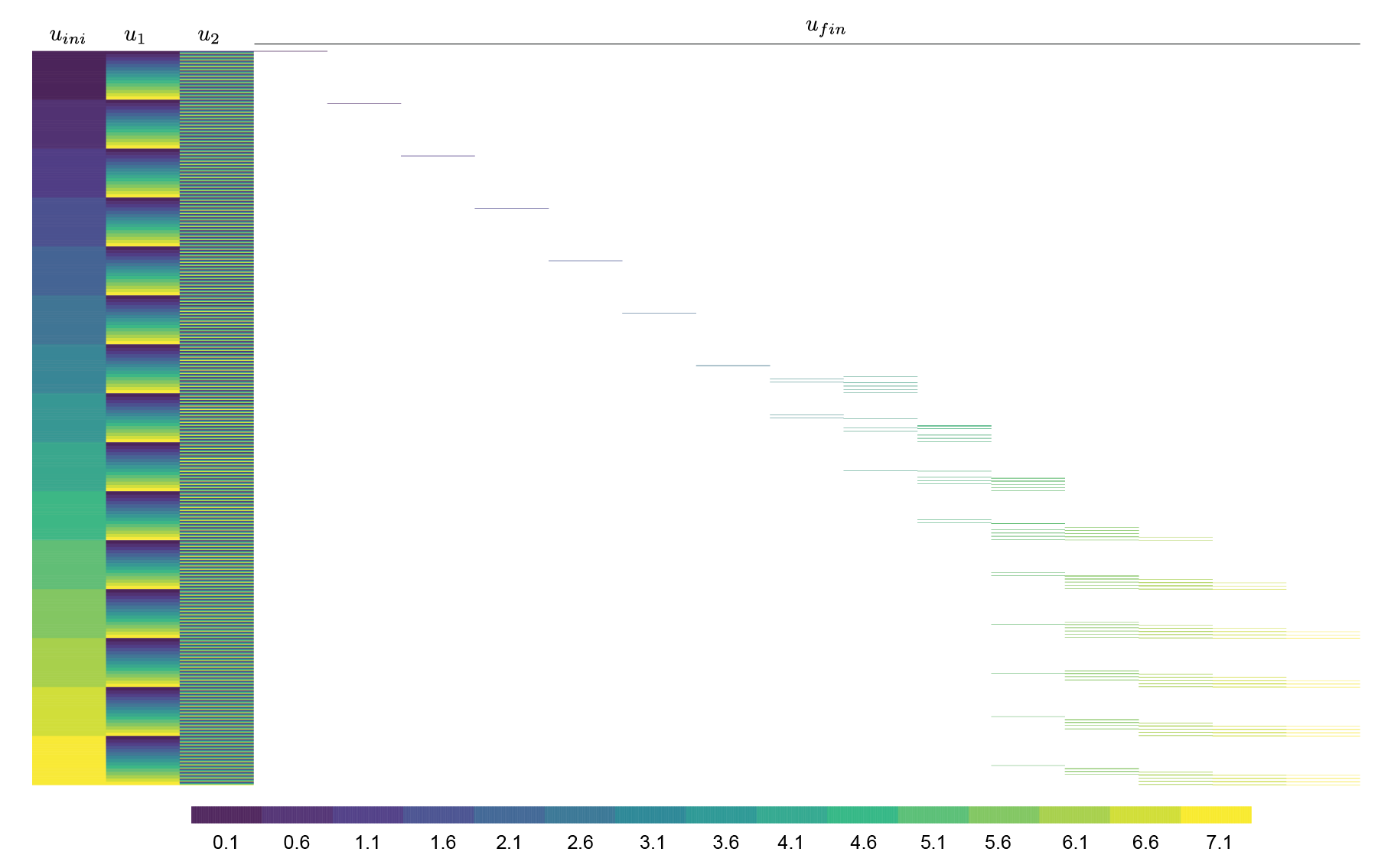}
    \caption{Complete list of all $n=2$ protocols. The first three columns from the left represent $u_{ini}, u_1, u_2$, respectively. We can denote the $u_{ini}, u_1, u_2$ as ``pre-protocol''. Each row represents one possible pre-protocol. If a pre-protocol $u_{ini}, u_1, u_2$ can be driven to the final equilibrium by any ultimate control $u_{fin}$, then the cell of this row at column corresponding to the value of $u_{fin}$ is filled by color. A row without filled cells on the right columns indicates that the corresponding pre-protocol can not be used by any hasty shortcut.}
    \label{fig:all_two_step_trajectories}
\end{figure*}

In Fig.~\ref{fig:all_two_step_trajectories}, we visualize the complete list of $15^3$ trajectories for $(n=2)$-step protocols sampled during the generation of the reachable set $R_s(2)$ and demonstrate that nontrivial shortcuts only exist for initial and final $c_s \geq 3.6$.
 
\begin{figure*}
    \centering
    \includegraphics[scale=1]{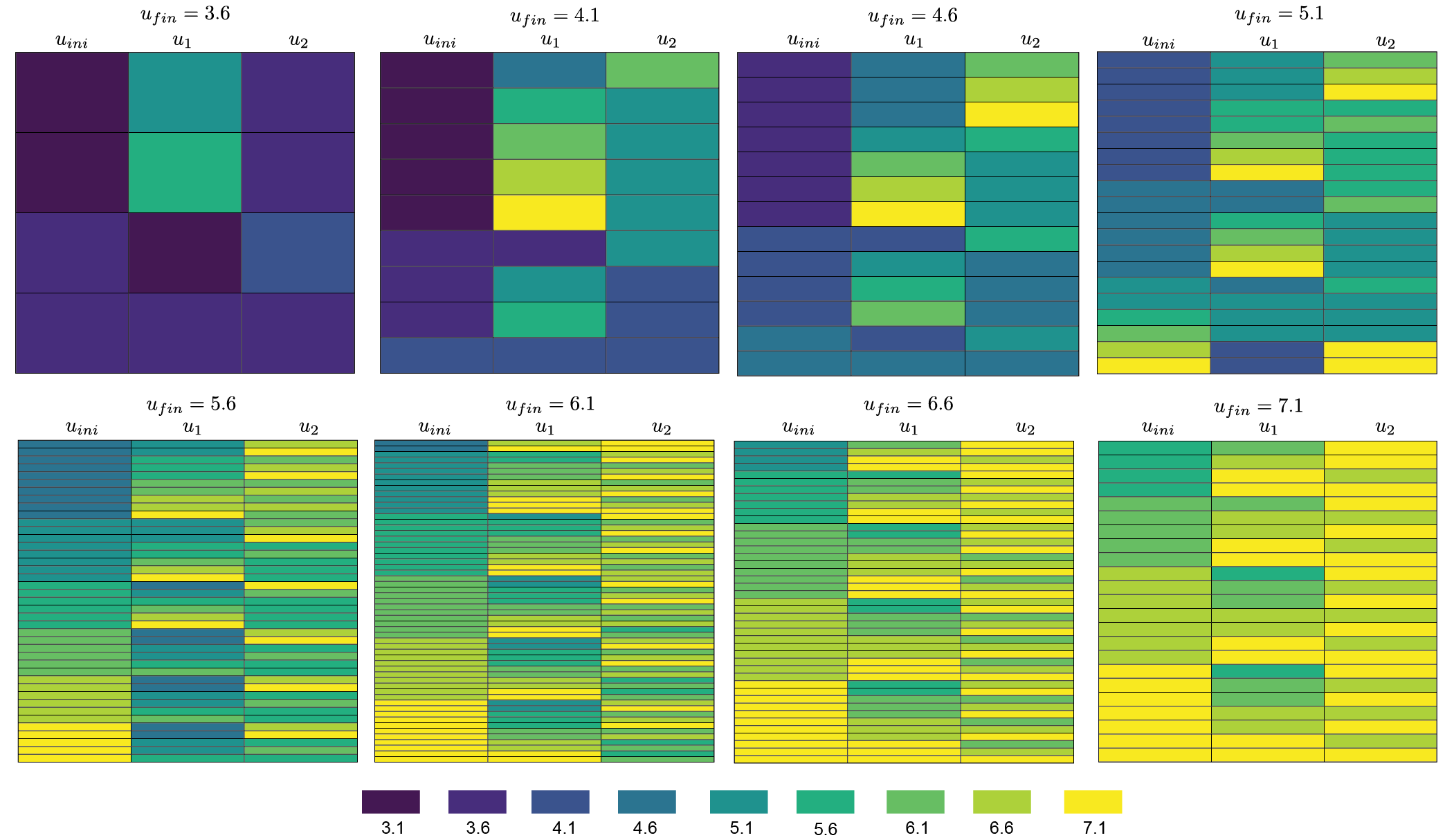}
    \caption{Pre-protocols ($u_{ini},u_1,u_2$) that can steer a system to $\vec p^{~ss}_{u_{fin}}$ by a final control $u_{fin}$ are grouped by the values of $u_{fin}$. The three columns represent $u_{ini}, u_1, u_2$, respectively. The header above each heatmap represents the target final distribution $u_{fin}$.}
    \label{fig:two_step_classes}
\end{figure*}

In Fig.~\ref{fig:two_step_classes}, shown are the protocols $u_{ini},u_1,u_2$ that can lead to a shortcut when combined with any $u_{fin}= c_s \in \{ 3.6, \cdots, 7.1\}$. They are grouped by different values of $u_{fin}$.

\begin{figure*}
    \centering
    \includegraphics[scale=1]{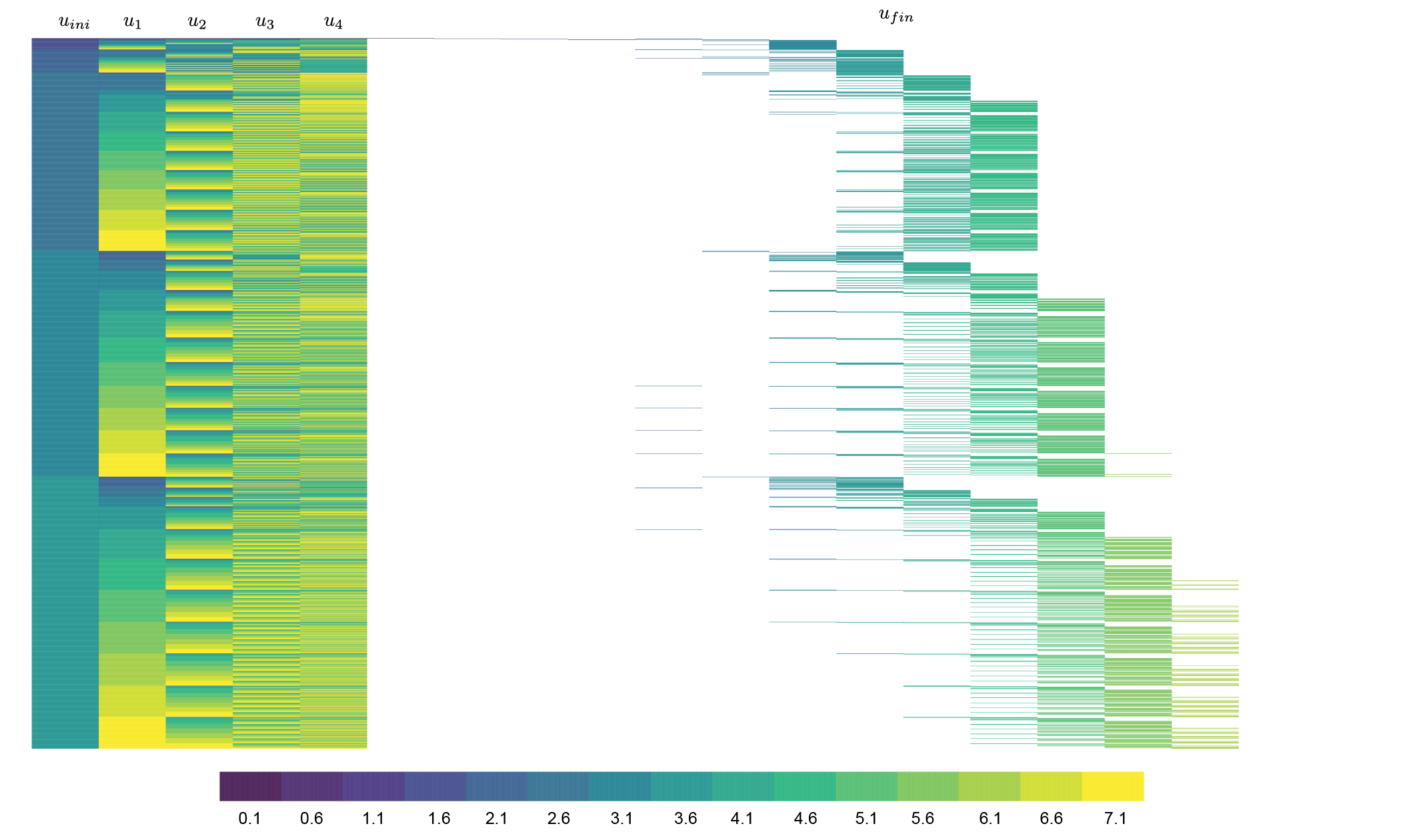}
    \caption{Complete list of all $n=4$ protocols. The first five columns from the left represent $u_{ini}, u_1, u_2,u_3,u_4$, respectively. We can denote the $u_{ini}, u_1, u_2,u_3,u_4$ as ``pre-protocol''. Each row represents one possible pre-protocol. If a pre-protocol $u_{ini}, u_1, u_2,u_3,u_4$ can be driven to the final equilibrium by any ultimate control $u_{fin}$, then the cell of this row at column corresponding to the value of $u_{fin}$ is filled by color. A row without filled cells on the right columns indicates that the corresponding pre-protocol can not be used by any hasty shortcut.}
    \label{fig:four_step_trajectories}
\end{figure*}

In Fig.~\ref{fig:four_step_trajectories}, we demonstrates the temporal protocls of $n=4$ for the low concentration range, $u_{ini} = 0.1, \cdots, 3.6$. Hasty shortcuts that are reachable to any $u_{fin}$ can be found by identifying the rows with filled color cells for the given column ($u_{fin}$).

\begin{figure*}
    \centering
    \includegraphics{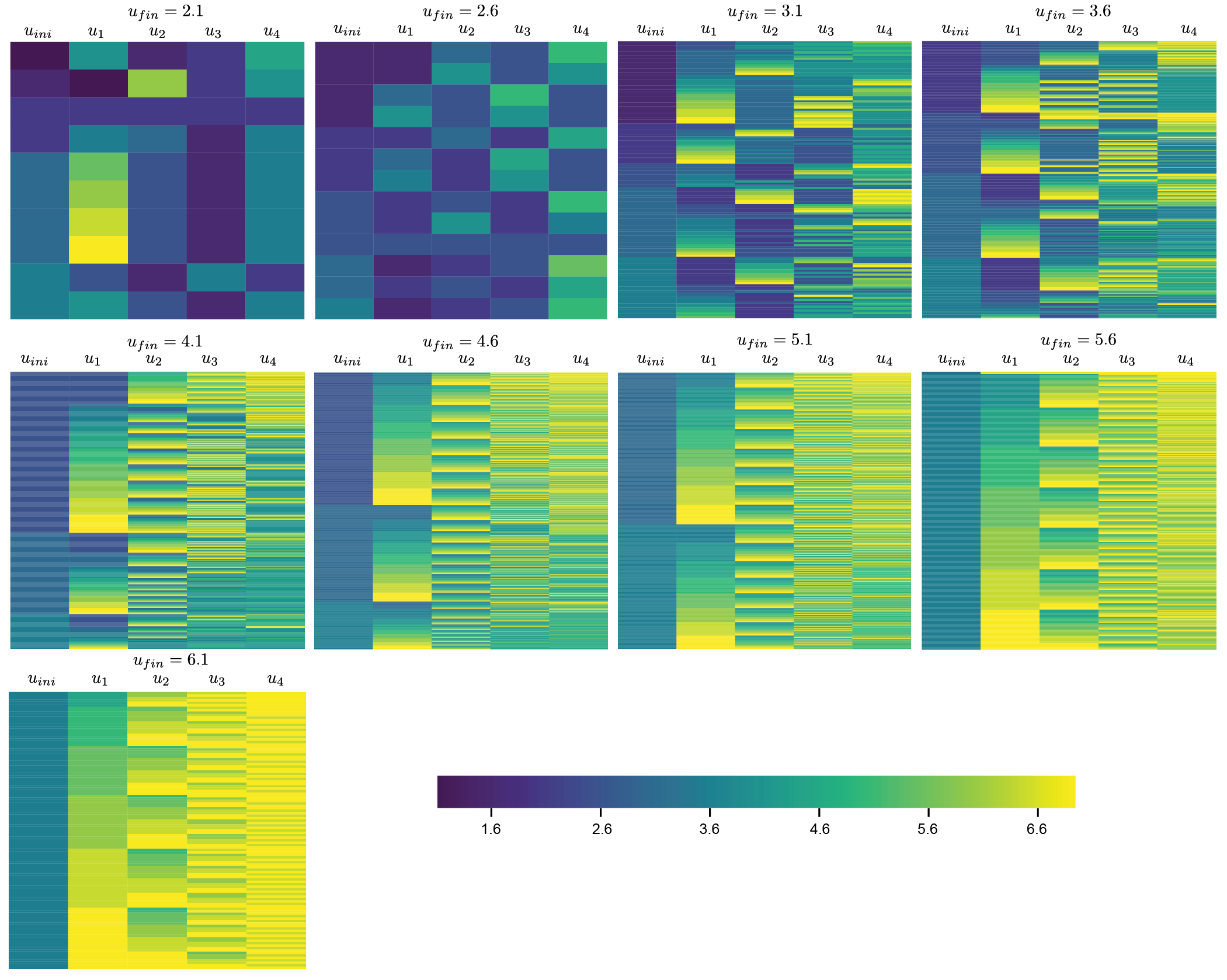}
    \caption{Pre-protocols ($u_{ini},u_1,u_2, u_3, u_4$) that can steer a system to $\vec p^{~ss}_{u_{fin}}$ by a final control $u_{fin}$ are grouped by the values of $u_{fin}$. The five columns represent $u_{ini}, u_1, u_2, u_3, u_4$, respectively. The header above each heatmap represents the target final distribution $u_{fin}$.}
    \label{fig:four_step_classes}
\end{figure*}

In Fig.~\ref{fig:four_step_classes}, hasty shortcut control protocols that can arrive at $u_{fin} \rightarrow c_s = 2.1, \cdots 6.1$ are grouped by the value of $u_{fin}$.

\section{Codes}
The numerical code written in Julia language will be made publicly available at GitHub.

\bibliography{apssamp}

\end{document}